\documentclass[11pt,draftclsnofoot,peerreviewca,onecolumn]{IEEEtran}
\usepackage{graphicx,amsmath,amssymb}
\pdfoutput=1
\begin{document}
%\begin{center}
% \title{\textbf{Technical Report on: Bayesian CRLB for Joint AoA, AoD and Multipath Gain Estimation in Millimeter Wave Wireless Networks}}
% \author{Laxminarayana S Pillutla and Ramesh Annavajjala}
% % (Date: February 08, 2015)\\ (Exam Duration: 90 minutes)}
% \date{March 2017}
%\end{center}
% \begin{titlepage}
% 
%                 \begin{center}
% 
%             \title{Technical Report on: Bayesian CRLB for Joint AoA, AoD and Multipath Gain Estimation in Millimeter Wave Wireless Networks}
%             \author{Laxminarayana S Pillutla and Ramesh Annavajjala}
%             \date{March 2017}
%     
%         \end{center}
% 
%     \end{titlepage}
% \begin{titlepage}
%     \centering
%     \vfill
%     {\bfseries\Large Technical Report on: Bayesian CRLB for Joint AoA, AoD and Multipath Gain Estimation in Millimeter Wave Wireless Networks\\
%         March 2017\\
%         \vskip2cm
%         Authors: Laxminarayana S Pillutla and Ramesh Annavajjala\\
%     }    
%     \vfill
%     \includegraphics[width=4cm]{daiict_logo.png} % also works with logo.pdf
%     \vfill
%     \vfill
% \end{titlepage}
\newtheorem{theorem}{\bf{Theorem}}
\newtheorem{proposition}{\bf{Proposition}}
\newtheorem{proof}{\bf{Proof}}
\title{CRLB Calculations for Joint AoA, AoD and Multipath Gain Estimation in Millimeter Wave Wireless Networks}
\date{April, 2017}
\author{Laxminarayana S Pillutla and Ramesh Annavajjala
\thanks{This technical report provides detailed calculations with respect to the derivation of non-random and random CRLB for the joint estimation problem of AoA, AoD and multipath gains.}
\thanks{The first author is with the Dhirubhai Ambani Institute of Information and Communication Technology (DA-IICT), Gandhinagar, Gujarat, India and the second author is with the NorthEastern University, Boston, MA, USA.
E-mail:\{\texttt{laxminarayana\_pillutla@daiict.ac.in, ramesh.annavajjala@gmail.com}\}}
%\thanks{$^{*}$ Corresponding Author. Address:
%Room \# 4203, Faculty Block - 4
%Dhirubhai Ambani Institute of Information and Communication Technology (DA-IICT), Post Bag 4 (Near Indroda Circle), Gandhinagar, Gujarat - 382007.
%Telephone: +91 79 30510621.
%}
}
\maketitle
\begin{abstract}
% In this letter we consider the derivation of non-random and Bayesian CRLB for the joint angle-of-arrival (AoA) and angle-of-departure (AoD) estimation problem. The expression for Bayesian CRLB is amenable to computations for various, AoA and AoD distributions and fading distributions, as long as the AoA, AoD and fading process on various paths are independent of each other. Numerical results based on the derived Bayesian CRLB, assuming AoA and AoD follow uniform distribution and fading on various paths to be Rician, suggest that the CRLB decreases with an increase in the Rice factor. Further we also explore the variations in CRLB for different types of beam forming and combining code books.
%\section{Abstract}
In this report we present an analysis of the non-random and the Bayesian Cramer-Rao lower bound (CRLB) for the joint estimation of angle-of-arrival (AoA), angle-of-departure (AoD), and the multipath amplitudes, for the millimeter-wave (mmWave) wireless networks. Our analysis is applicable to multipath channels with Gaussian noise and independent path parameters. Numerical results based on uniform AoA and AoD in $[0,\pi)$, and Rician fading path amplitudes, reveal that the Bayesian CRLB decreases monotonically with an increase in the Rice factor. Further, the CRLB obtained by using beamforming and combining code books generated by quantizing directly the domain of AoA and AoD was found to be lower than those obtained with other types of beamforming and combining code books.
\end{abstract}
\IEEEpeerreviewmaketitle
\section{Introduction}
Millimeter wave (mmWave) communications are explored to push data rates in wireless networks to the order of gigabit per second. Directional transmissions would be used in mmWave networks to counter the increased path loss at mmWave frequencies ($6$ GHz and beyond) \cite{RHDM2014}. This requires accurate beam alignment on either side of the communication link, a task in turn requires accurate estimation of angle-of-departure (AoD) and angle-of-arrival (AoA) at the transmitter and receiver respectively. In this report we consider the fundamental performance limits of joint AoD and AoA estimation, in terms of the well known Cramer-Rao lower bound (CRLB) computation.

The problem of CRLB computation for AoA estimation alone was previously considered in \cite{SN89}. However AoD estimation and the signal fading effect (that is inherent in any wireless communication scenario) were not considered in \cite{SN89}. The work in \cite{SN89} further analyzes the statistical efficiency of MUSIC and maximum-likelihood (ML) methods. Other works that consider AoA estimation and CRLB computation can be found in \cite{CWG_springer_2016} and \cite{FND_ieee_taes}. In \cite{arxiv_1702_00276}, Bayesian CRLB was derived for efficient beam tracking in mmWave communications, which deals with estimation/tracking of AoD only. The joint AoA and AoD estimation problem and the concomitant channel estimation problem for mmWave communications was considered in several recent works like those in \cite{AALH2014_jstp} and \cite{ZGF_icc2016}, to name a few. However, these works do not deal with CRLB computations. 

In this report we consider both the non-random and Bayesian CRLB derivations for the joint AoA and AoD estimation problem. The derived Bayesian CRLB would serve as a ground truth against which the performance of any unbiased estimator can be compared. The interested readers are referred to the monograph in \cite{Trees:2007:BBP:1296178} for an exhaustive collection of works on Bayesian bounds for parameter estimation. We consider a system model that uses pilot symbols at the transmitter and receiver to facilitate the estimation of AoA and AoD at the receiver. Our specific contributions for the system model we assumed are as follows:
\begin{enumerate}
\item We first derive the non-random CRLB, for the observation model that we assumed, which any unbiased estimator for AoA and AoD should satisfy (cf. Theorem \ref{theorem:nr_crlb} in Section \ref{subsect:nr_crlb}).
\item We next derive the Bayesian CRLB assuming (i) the AoA and AoD distribution to be uniform (although the underlying calculations are equally suitable for other distributions) and (ii) the signal fading process on various paths between the transmitter and receiver to be of Rician distribution (which includes Rayleigh fading and AWGN as special cases) (cf. Proposition \ref{prop:b_crlb_sp_case} in Section \ref{subsect:bayesian_crlb}).
\item Our numerical results obtained based on the derived Bayesian CRLB point to the following important conclusions: (i) An increase in the Rice factor causes the Bayesian CRLB to decrease and (ii) beam forming and combining code books generated by directly quantizing the domain of AoA and AoD ($=[0,\pi)$) resulted in a lower CRLB value compared to the cases, in which beam forming and combining code books are generated by quantizing the range of AoA and AoD ($=[-1,1]$) and the one in which orthogonal beam forming and combining code books are used.
\end{enumerate}
% The rest of this letter is organized as follows: Section \ref{sect:system_model} contains a system model description of the channel estimation problem considered in this paper. The Section \ref{sect:crlb} contains derivations for CRLB and Bayesian CRLB expressions, followed by computation of Bayesian CRLB for the special case when angle of departure at the transmitter and angle of arrival at the receiver follow a uniform distribution. The Section \ref{sect:numresults} contains numerical results of derived CRLB and simulation results of the proposed algorithm and Section \ref{sect:conclusions} contains important conclusions of this work.
\subsection*{Notation}
Throughout this report we use boldface capital alphabets to denote matrices and boldface small alphabets to denote the vectors. A diagonal matrix is denoted as $\text{diag}(\mathbf{q})$ such that its diagonal elements are equal to the elements of $\mathbf{q}$. The symbols $*$, $T$ and $\dagger$ are used to denote complex conjugate, matrix transpose and matrix conjugate transpose operators respectively. The Kronecker product between two matrices $\mathbf{A}$ and $\mathbf{B}$ is denoted as $\mathbf{A} \otimes \mathbf{B}$. The mathematical expectation operation is denoted as $\mathbb{E}[.]$.
\section{System Model Description of Channel Estimation in Millimeter Wave Wireless Systems}
\label{sect:system_model}
\subsection{System and channel model description}
\label{subsect:system_model_1}
Consider a typical mmWave communication system which comprises of a transmitter and receiver equipped with uniform linear arrays (ULA) with antenna elements $n_t$ and $n_r$ respectively. The spacing between antenna elements is assumed to be equal to half of the corresponding radiated signal's wavelength. Denote $\phi$ and $\psi$ as the angle of arrival (AoD) and angle of arrival (AoA) at the ULA of transmitter and receiver respectively. The corresponding beamsteering vectors are then given by $\mathbf{e}_{t}(\phi)=\frac{1}{\sqrt{n_t}}\left[1,e^{-j \pi \cos(\phi)},\cdots,e^{-j\pi(n_{t}-1) \cos(\phi)}\right]^{T}$ and $\mathbf{e}_{r}(\phi)=\frac{1}{\sqrt{n_r}}\left[1,e^{-j \pi \cos(\psi)},\cdots,e^{-j\pi(n_{r}-1) \cos(\psi)}\right]^{T}$ respectively. Therefore under the $L$-path scatterer model the channel matrix $\mathbf{H}$ is given by
\begin{equation}
\mathbf{H}=\sum_{l=1}^{L}\alpha_{l}\mathbf{e}_{r}(\psi_{l})\mathbf{e}_{t}(\phi_{l})^{\dagger},\label{channel_matrix_def}
\end{equation}
where $\alpha_{l} > 0~(l=1,\cdots,L)$ denote the random path gains that are assumed to be Ricean distributed with the normalized probability density function (PDF) \cite[Eq. (2.3-61)]{PS2008} and $\phi_l$ and $\psi_{l}$ (which denote AoD and AoA along path $l$) are assumed to be uniformly distributed between $0$ and $\pi$. The CRLB calculations in Section \ref{sect:crlb} are equally applicable for other statistical distributions of $\alpha_{l}, \phi_{l}$ and $\psi_{l}$, so long as they are mutually independent of each other and also independent of the corresponding parameters across other paths.

%$2\left(K + 1\right)xe^{-\left(K + 1\right)\left(x^{2} + \frac{K}{(K + 1)}\right)}I_{0}\left(2 x\sqrt{K(K+1)}\right)u(x)$ ($K$ i denotes the Rice factor and $u(x)$ denotes the unit step function)
\subsection{Pilot based channel estimation}
In this subsection we consider a pilot based method for estimation of the channel gain, AoD and AoA across various paths at the receiver. The values thus estimated can be used for re-construction of channel matrix at the receiver.

Let $p_t$ and $p_r$ denote the number of beamforming and beamcombining pilot symbols used to facilitate estimation of path gain, AoA and AoD, associated with various paths at the receiver. The beamforming and beamcombining vectors, associated with the directions $\tilde{\phi}_{p}~(p=1,\cdots, p_{t})$ and $\tilde{\psi}_{p}~(p=1,\cdots, p_{r})$, are given by $\mathbf{f}_{p} \triangleq \mathbf{e}_{t}(\tilde{\phi}_{p})$ and $\mathbf{g}_{p} \triangleq \mathbf{e}_{r}(\tilde{\psi}_{p})$ respectively (Note: $\mathbf{e}_{t}(.)$ and $\mathbf{e}_{r}(.)$ are as defined in Section \ref{subsect:system_model_1}). The $p^{\text{th}}$ beamforming pilot vector $\mathbf{f}_{p}$ transmitted, passes through the channel with matrix defined in Eq. (\ref{channel_matrix_def}) and is combined at the receiver using the $q^{\text{th}}$ beamcombining pilot vector $\mathbf{g}_{q}$. The resulting noisy observation is given by $y_{pq}=\mathbf{g}_{q}^{\dagger} \mathbf{H} \mathbf{f}_{p} + \mathbf{g}_{q}^{\dagger} \mathbf{z}$, where $\mathbf{z}$ denotes the received complex noise vector that is assumed to be proper and Gaussian distributed with mean $\mathbf{0}$ and covariance matrix $\sigma_{v}^{2}I_{n_{r}}$. If each of the $p_t$ beamforming pilots are repeated $p_r$ times, then they can be combined using the $p_r$ beamcombining pilots at the receiver and the resulting $p_t p_r$ observations can be expressed in matrix form using matrices $\mathbf{F} \triangleq \left[\mathbf{f}_{1}, \mathbf{f}_{2},\cdots,\mathbf{f}_{p_{t}}\right]$ and $\mathbf{G} \triangleq \left[\mathbf{g}_{1},\mathbf{g}_{2},\cdots,\mathbf{g}_{p_{r}}\right]$ as
\begin{eqnarray}
&&\mathbf{Y}=\mathbf{G}^{\dagger} \mathbf{H} \mathbf{F} + \mathbf{Z}~,\text{Define}~\mathbf{y} \triangleq \text{\textbf{vec}}\left(\mathbf{Y}\right)\label{observation_matrix_def}\\
&&\mathbf{y}=\sum_{l=1}^{L} \left(\underbrace{\mathbf{F}^{T} \otimes \mathbf{G}^{\dagger}}_{\mathbf{A}}\right)\left(\mathbf{e}_{t}^{*}(\phi_{l}) \otimes  \mathbf{e}_{r}(\psi_{l})\right)+ \mathbf{v}~,\label{observation_vector_def}
\end{eqnarray}
where $\mathbf{Z}$ denotes a complex matrix whose entries are assumed to be uncorrelated and proper, each of zero mean and variance $\sigma_{v}^{2}$ and $\mathbf{v}=\text{\textbf{vec}}(\mathbf{Z})$ (which is a proper complex Gaussian random vector of mean $\mathbf{0}$ and covariance matrix $\sigma_{v}^{2}I_{p_{t} p_{r}}$). The Eq. (\ref{observation_vector_def}) is obtained by using Eq. (\ref{channel_matrix_def}) and the following property: $\text{\textbf{vec}}(\mathbf{UVW}) = \left(\mathbf{W}^{T} \otimes \mathbf{U}\right) \text{\textbf{vec}}(\mathbf{V})$, where $\mathbf{U}$, $\mathbf{V}$ and $\mathbf{W}$ are matrices of suitable dimensions \cite[Appendix A]{vantrees_part4}.
\section{Cramer-Rao Lower Bound Expression for Unbiased Channel Estimator}
\label{sect:crlb}
\subsection{Log-likelihood ratio of the given observation model}
\label{subsect:crlb_llr_def}
For the purpose of analysis in this section we define the following:
\begin{eqnarray*}
\boldsymbol{\theta}&\triangleq&\left[\phi_{1},\cdots,\phi_{L},\psi_{1},\cdots,\psi_{L},\alpha_{1},\cdots,\alpha_{L}\right]^{T}\\,
\tilde{\mathbf{e}}_{t}(\phi_{l}) &\triangleq& \frac{1}{\sqrt{n_t}}\left[0,e^{j \pi \cos(\phi_{l})},\cdots,(n_{t}-1)e^{j (n_{t} -1) \pi \cos(\phi_{l})}\right]^{T},\\ 
\tilde{\mathbf{e}}_{r}(\psi_l) &=& \frac{1}{\sqrt{n_r}}\left[0,e^{j \pi \cos(\psi_{l})},\cdots,(n_{r}-1)e^{j (n_{r}-1)\pi \cos(\psi_{l})}\right]^{T}\\
\mathbf{m} &\triangleq& \sum_{l=1}^{L} \mathbf{A} \left(\mathbf{e}_{t}^{*}(\phi_{l}) \otimes  \mathbf{e}_{r}(\psi_{l})\right)~,\\
\text{and}~\mathbf{K} &\triangleq& \mathbf{A}^{\dagger} \mathbf{A}.
\end{eqnarray*}

The model likelihood function $p\left(\mathbf{y}|\boldsymbol{\theta}\right)$ is given by the PDF of a proper complex Gaussian random vector with mean vector $\mathbf{m}$ and covariance matrix $\sigma_{v}^{2}I_{m_{r}m_{t}}$, which is equal to $\frac{1}{\pi^{p_{r}p_{t}} (\sigma_{v}^{2})^{p_{r}p_{t}}} e^{-\frac{1}{\sigma_{v}^{2}}\left(\mathbf{y} - \mathbf{m}\right)^{\dagger}\left(\mathbf{y} - \mathbf{m}\right)}$ \cite{PS2008}. Therefore, the log-likelihood function of the model is given by
%and its first derivative with respect to any component of $\boldsymbol{\theta}$, say $\theta_{i}$ (to be used in the FIM computation) are given by
\begin{eqnarray}
&&L(\mathbf{y})=-p_{r}p_{t}\ln(\pi \sigma_{v}^{2}) - \frac{1}{\sigma_{v}^{2}}\left[\left(\mathbf{y} - \mathbf{m}\right)^{\dagger}\left(\mathbf{y} - \mathbf{m}\right)\right]~.\label{llr_eqn}
%&&\frac{\partial L(\mathbf{y})}{\partial \theta_{i}}=\frac{1}{\sigma_{v}^{2}}\left[\left(\mathbf{y}^{\dagger} - \mathbf{m}^{\dagger}\right)\frac{\partial \mathbf{m}}{\partial \theta_{i}} + \frac{\partial \mathbf{m}^{\dagger}}{\partial \theta_{i}}\left(\mathbf{y} - \mathbf{m}\right)\right]\label{llr_first_der}
\end{eqnarray}
\subsection{FIM Computation and the non-random CRLB}
\label{subsect:nr_crlb}
Let $\hat{\boldsymbol{\theta}}$ denote the unbiased estimator of $\boldsymbol{\theta}$, then the CRLB on variance of $\hat{\boldsymbol{\theta}}$ is given by $\text{Var}\left(\hat{\boldsymbol{\theta}}\right) \geq \text{tr}\left(\widetilde{\mathbf{J}}_{\text{NR}}^{-1}\right)$, where $\widetilde{\mathbf{J}}_{\text{NR}}$ denotes the FIM whose $(i,j)^{\text{th}}$ entry is given by $\widetilde{\mathbf{J}}_{\text{NR}}(i,j) \triangleq -\mathbb{E}\left[\frac{\partial^{2} L(\mathbf{y})}{\partial \theta_{i} \partial \theta_{j}}\right]$ ($\mathbb{E}$ w.r.t the PDF of $(\mathbf{y}|\boldsymbol{\theta})$). The CRLB of any unbiased estimator of $\boldsymbol{\theta}$ defined as in Section \ref{subsect:crlb_llr_def} is given by the following theorem.
\begin{theorem}
For the observation model defined in Eq. (\ref{observation_vector_def}) the non-random CRLB for any unbiased estimator of $\boldsymbol{\theta}$ is given by $\text{tr}(\mathbf{J}_{\text{NR}}^{-1})$. The entries of matrix $\mathbf{J}_{\text{NR}}$ are given as follows: (the term corresponding to $-\mathbb{E}\left[\frac{\partial^{2} L(\mathbf{y})}{\partial \theta_{l} \partial \theta_{m}}\right]$ in the matrix $\mathbf{J}_{\text{NR}}$ is denoted as $\mathbf{J}_{\text{NR}}\left(\theta_{l},\theta_{m}\right)$): 
\begin{eqnarray}
\label{nr_fim_eq}
\begin{split}
&\mathbf{J}_{\text{NR}}(\phi_{l},\phi_{l})=\frac{2 \pi^2 \alpha_{l}^{2}}{\sigma_{v}^{2}}\text{tr}\left[\mathbf{K}\left(\mathbf{P} \otimes \mathbf{Q}\right)\right]\\
&\mathbf{J}_{\text{NR}}(\phi_{l},\alpha_{l})=\frac{-2 \pi \alpha_{l}}{\sigma_{v}^{2}} \text{Im}\left\{\text{tr}\left[\mathbf{K} \left(\mathbf{P}^{(2)} \otimes \mathbf{Q}\right) \right]\right\}\\
&\mathbf{J}_{\text{NR}}(\phi_{l},\psi_{l})=\frac{-2 \pi^{2}\alpha_{l}^{2}}{\sigma_{v}^{2}} \text{Re}\left\{\text{tr}\left[\mathbf{K} \left(\mathbf{P}^{(3)} \otimes \mathbf{Q}^{(2)}\right) \right]\right\}\\ &\mathbf{J}_{\text{NR}}(\psi_{l},\alpha_{l})=\frac{2 \pi \alpha_{l}}{\sigma_{v}^{2}}\text{Im}\left\{\text{tr}\left[\mathbf{K} \left(\mathbf{P}^{(4)} \otimes \mathbf{Q}^{(2)}\right)\right]\right\}\\
&\mathbf{J}_{\text{NR}}(\phi_{l},\alpha_{m})=\frac{-2 \pi \alpha_{l}}{\sigma_{v}^{2}}\text{Im}\left\{\text{tr}\left[\mathbf{K} \left(\mathbf{P}^{(5)} \otimes \mathbf{Q}^{(3)}\right)\right]\right\}\\
&\mathbf{J}_{\text{NR}}(\psi_{l},\alpha_{m})=\frac{2 \pi \alpha_{l}}{\sigma_{v}^{2}}\text{Im}\left\{\text{tr}\left[\mathbf{K} \left(\mathbf{P}^{(6)} \otimes \mathbf{Q}^{(4)}\right)\right]\right\}\\
&\mathbf{J}_{\text{NR}}(\phi_{l},\phi_{m})=\frac{2 \pi^{2} \alpha_{l} \alpha_{m}}{\sigma_{v}^{2}}\text{Re}\left\{\text{tr}\left[\mathbf{K} \left(\mathbf{P}^{(7)} \otimes \mathbf{Q}^{(3)}\right)\right]\right\}\\
&\mathbf{J}_{\text{NR}}(\psi_{l},\psi_{m})=\frac{2 \pi^{2} \alpha_{l}\alpha_{m}}{\sigma_{v}^{2}}\text{Re}\left\{\text{tr}\left[\mathbf{K} \left(\mathbf{P}^{(8)} \otimes \mathbf{Q}^{(5)}\right)\right]\right\}\\
&\mathbf{J}_{\text{NR}}(\psi_{l},\psi_{l})=\frac{2 \pi^2 \alpha_{l}^{2}}{\sigma_{v}^{2}}\text{tr}\left(\mathbf{K} \left(\mathbf{P}^{(4)} \otimes \mathbf{Q}^{(6)}\right)\right) \\
&\mathbf{J}_{\text{NR}}(\phi_{l},\psi_{m})=\frac{-2 \pi^{2}\alpha_{l}\alpha_{m}}{\sigma_{v}^{2}}\text{Re}\left\{\text{tr}\left[\mathbf{K} \left(\mathbf{P}^{(9)} \otimes \mathbf{Q}^{(7)}\right)\right]\right\}\\
&\mathbf{J}_{\text{NR}}(\alpha_{l},\alpha_{l})=\frac{2}{\sigma_{v}^{2}}\text{tr}\left[\mathbf{K} \left(\mathbf{P}^{(4)} \otimes \mathbf{Q}\right)\right]\\
&\mathbf{J}_{\text{NR}}(\alpha_{l},\alpha_{m})=\frac{2}{\sigma_{v}^{2}}\text{Re}\left\{\text{tr}\left[\mathbf{K} \left(\mathbf{P}^{(6)} \otimes \mathbf{Q}^{(3)}\right)\right]\right\}~.
\end{split}
\end{eqnarray}
The matrices $\mathbf{P}, \cdots, \mathbf{P}^{(9)}$ and $\mathbf{Q}, \cdots, \mathbf{Q}^{(7)}$ in the Eq. (\ref{nr_fim_eq}) above are defined and computed in Appendix A.
\newline
\textbf{Proof:} See Appendix A.
\label{theorem:nr_crlb}
\end{theorem}
%where P \triangleq \tilde{\mathbf{e}}_{t}(\phi_{l}) \tilde{\mathbf{e}}^{\dagger}_{t}(\phi_{l})\sin^{2}(\phi_{l})
% \begin{proof}
% See Appendix A.
% \end{proof}
\subsection{Bayesian FIM Computation and the Bayesian CRLB}
\label{subsect:bayesian_crlb}
The non-random CRLB in Theorem \ref{theorem:nr_crlb} depends on specific realization of the parameter $\boldsymbol{\theta}$. However, in many cases of interest one desires for a lower bound on MSE of any estimator that is independent of the estimated parameter realization. For such cases one can derive the Bayesian CRLB, which is given by the $\text{tr}(\widetilde{\mathbf{J}}_{\text{B}}^{-1})$, where $\widetilde{\mathbf{J}}_{\text{B}}$ denotes the Bayesian FIM. The Bayesian FIM can be decomposed into two components i.e. $\widetilde{\mathbf{J}}_{\text{B}}=\widetilde{\mathbf{J}}_{\text{D}} + \widetilde{\mathbf{J}}_{\text{P}}$, where $\widetilde{\mathbf{J}}_{\text{D}}$ denotes the part corresponding to the (observation) data and $\widetilde{\mathbf{J}}_{\text{P}}$ denotes the path corresponding to the prior knowledge \cite{vantrees_part4}. By definition $\widetilde{\mathbf{J}}_{\text{D}}(i,j) \triangleq -\mathbb{E}\left[\frac{\partial^{2} L\left(\mathbf{y}\right)}{\partial \theta_{i} \theta_{j}}\right]$, $\mathbb{E}$ is w.r.t the joint PDF of $(\mathbf{y},\boldsymbol{\theta})$. By definition $\widetilde{\mathbf{J}}_{\text{P}} (i,j) \triangleq -\mathbb{E}\left[\frac{\partial^{2} \ln(p(\boldsymbol{\theta}))}{\partial \theta_{i} \theta_{j}}\right]$, $\mathbb{E}$ is w.r.t the PDF of $\boldsymbol{\theta}$ \cite{vantrees_part4}. In our case the matrix $\widetilde{\mathbf{J}}_{\text{D}}$ is obtained by simply averaging $\mathbf{J}_{\text{NR}}$ w.r.t the PDF of $\boldsymbol{\theta}$ (since averaging over the PDF of $(\mathbf{y}|\boldsymbol{\theta})$ was already done during the computation of $\mathbf{J}_{\text{NR}}$). In averaging the various entries of $\mathbf{J}_{\text{NR}}$ w.r.t the PDF of $\boldsymbol{\theta}$, it is worthwhile to note that the gains, AoA and AoD across various paths are independent of each other. Further for a given path the multipath gain, AoA and AoD are independent of each other. Similarly $\widetilde{\mathbf{J}}_{\text{P}} (i,j) \triangleq -\mathbb{E}\left[\frac{\partial^{2} \ln(p(\boldsymbol{\theta}))}{\partial \theta_{i} \theta_{j}}\right]$, $\mathbb{E}$ is w.r.t the PDF of $\boldsymbol{\theta}$ which is denoted as $p(\boldsymbol{\theta})$. 
%It is a straighforward exercise to show that all the terms are equal to $0$, except the following $L$ terms given by $-\mathbb{E}\left[\frac{\partial^{2} \ln(p(\boldsymbol{\theta}))}{\partial \theta_{i}^{2}}\right]~(i=1,\cdots,L)$.

For the special case when AoDs and AoAs of all paths (i.e. $\{\phi_{l}\}_{l=1}^{L}$ and $\{\psi_{l}\}_{l=1}^{L}$) are uniformly distributed between $0$ and $\pi$, as we assumed in this report, the Bayesian FIM $\mathbf{J}_{\text{B}}$ and hence CRLB can be computed as in the following proposition.
\begin{proposition}
For the special case when $\phi_{l}, \psi_{l}~(l=1,\cdots,L)$ are uniformly distributed over $[0,\pi)$ the Bayesian CRLB for any unbiased estimator of $\boldsymbol{\theta}$ is given by $\text{tr}\left(\mathbf{J}_{\text{B}}^{-1}\right)$, where the matrix $\mathbf{J}_{\text{B}}= \mathbf{J}_{\text{D}} + \mathbf{J}_{\text{P}}$. The entries of $\mathbf{J}_{\text{P}}$ are all zero except for the following $L$ terms: $-\mathbb{E}\left[\frac{\partial^{2}\ln(p(\boldsymbol{\theta}))}{\partial \alpha_{l}^{2}}\right]~(l=1, \cdots, L)$, whose expressions can be obtained from Appendix B. 

The entries of $\mathbf{J}_{\text{D}}$ (similar to the entries of $\mathbf{J}_{\text{NR}}$) are given below:
\begin{eqnarray}
\label{r_d_fim_eq}
\begin{split}
&\mathbf{J}_{\text{NR}}(\phi_{l},\phi_{l})=\frac{2 \pi^2 \mathbb{E}[\alpha_{l}^{2}]}{\sigma_{v}^{2}}\text{tr}\left[\mathbf{K}\left(\widetilde{\mathbf{P}} \otimes \widetilde{\mathbf{Q}}\right)\right]\\ &\mathbf{J}_{\text{NR}}(\phi_{l},\alpha_{l})=\frac{-2 \pi \mathbb{E}[\alpha_{l}]}{\sigma_{v}^{2}} \text{Im}\left\{\text{tr}\left[\mathbf{K} \left(\widetilde{\mathbf{P}}^{(2)} \otimes \widetilde{\mathbf{Q}}\right) \right]\right\}\\
&\mathbf{J}_{\text{NR}}(\phi_{l},\psi_{l})=\frac{-2 \pi^{2}\mathbb{E}[\alpha_{l}^{2}]}{\sigma_{v}^{2}} \text{Re}\left\{\text{tr}\left[\mathbf{K} \left(\widetilde{\mathbf{P}}^{(3)} \otimes \widetilde{\mathbf{Q}}^{(2)}\right) \right]\right\}\\
&\mathbf{J}_{\text{NR}}(\psi_{l},\alpha_{l})=\frac{2 \pi \mathbb{E}[\alpha_{l}]}{\sigma_{v}^{2}}\text{Im}\left\{\text{tr}\left[\mathbf{K} \left(\widetilde{\mathbf{P}}^{(4)} \otimes \widetilde{\mathbf{Q}}^{(2)}\right)\right]\right\}\\
&\mathbf{J}_{\text{NR}}(\phi_{l},\alpha_{m})=\frac{-2 \pi \mathbb{E}[\alpha_{l}]}{\sigma_{v}^{2}}\text{Im}\left\{\text{tr}\left[\mathbf{K} \left(\widetilde{\mathbf{P}}^{(5)} \otimes \widetilde{\mathbf{Q}}^{(3)}\right)\right]\right\}\\
&\mathbf{J}_{\text{NR}}(\psi_{l},\alpha_{m})=\frac{2 \pi \mathbb{E}[\alpha_{l}]}{\sigma_{v}^{2}}\text{Im}\left\{\text{tr}\left[\mathbf{K} \left(\widetilde{\mathbf{P}}^{(6)} \otimes \widetilde{\mathbf{Q}}^{(4)}\right)\right]\right\}\\
&\mathbf{J}_{\text{NR}}(\phi_{l},\phi_{m})=\frac{2 \pi^{2} \mathbb{E}[\alpha_{l}] \mathbb{E}[\alpha_{m}]}{\sigma_{v}^{2}}\text{Re}\left\{\text{tr}\left[\mathbf{K} \left(\widetilde{\mathbf{P}}^{(7)} \otimes \widetilde{\mathbf{Q}}^{(3)}\right)\right]\right\}\\
&\mathbf{J}_{\text{NR}}(\psi_{l},\psi_{m})=\frac{2 \pi^{2} \mathbb{E}[\alpha_{l}]\mathbb{E}[\alpha_{m}]}{\sigma_{v}^{2}}\text{Re}\left\{\text{tr}\left[\mathbf{K} \left(\widetilde{\mathbf{P}}^{(8)} \otimes \widetilde{\mathbf{Q}}^{(5)}\right)\right]\right\}\\
&\mathbf{J}_{\text{NR}}(\psi_{l},\psi_{l})=\frac{2 \pi^2 \mathbb{E}[\alpha_{l}^{2}]}{\sigma_{v}^{2}}\text{tr}\left(\mathbf{K} \left(\widetilde{\mathbf{P}}^{(4)} \otimes \widetilde{\mathbf{Q}}^{(6)}\right)\right)\\
&\mathbf{J}_{\text{NR}}(\phi_{l},\psi_{m})=\frac{-2 \pi^{2}\mathbb{E}[\alpha_{l}]\mathbb{E}[\alpha_{m}]}{\sigma_{v}^{2}}\text{Re}\left\{\text{tr}\left[\mathbf{K} \left(\widetilde{\mathbf{P}}^{(9)} \otimes \widetilde{\mathbf{Q}}^{(7)}\right)\right]\right\}\\
&\mathbf{J}_{\text{NR}}(\alpha_{l},\alpha_{l})=\frac{2}{\sigma_{v}^{2}}\text{tr}\left[\mathbf{K} \left(\widetilde{\mathbf{P}}^{(4)} \otimes \widetilde{\mathbf{Q}}\right)\right]\\
&\mathbf{J}_{\text{NR}}(\alpha_{l},\alpha_{m})=\frac{2}{\sigma_{v}^{2}}\text{Re}\left\{\text{tr}\left[\mathbf{K} \left(\widetilde{\mathbf{P}}^{(6)} \otimes \widetilde{\mathbf{Q}}^{(3)}\right)\right]\right\}~.
\end{split}
\end{eqnarray}
The matrices $\widetilde{\mathbf{P}}, \cdots, \widetilde{\mathbf{P}}^{(9)}$ and $\widetilde{\mathbf{Q}}, \cdots, \widetilde{\mathbf{Q}}^{(7)}$ in Eq. (\ref{r_d_fim_eq}) are defined in Appendix C.
\newline
\textbf{Proof:} See Appendix B for the derivation of entries for $\mathbf{J}_{\text{P}}$ and Appendix C for the derivation of entries for $\mathbf{J}_{\text{D}}$.
\label{prop:b_crlb_sp_case}
\end{proposition}

\begin{figure}[p]
\begin{center}
\includegraphics[width=0.95\linewidth]{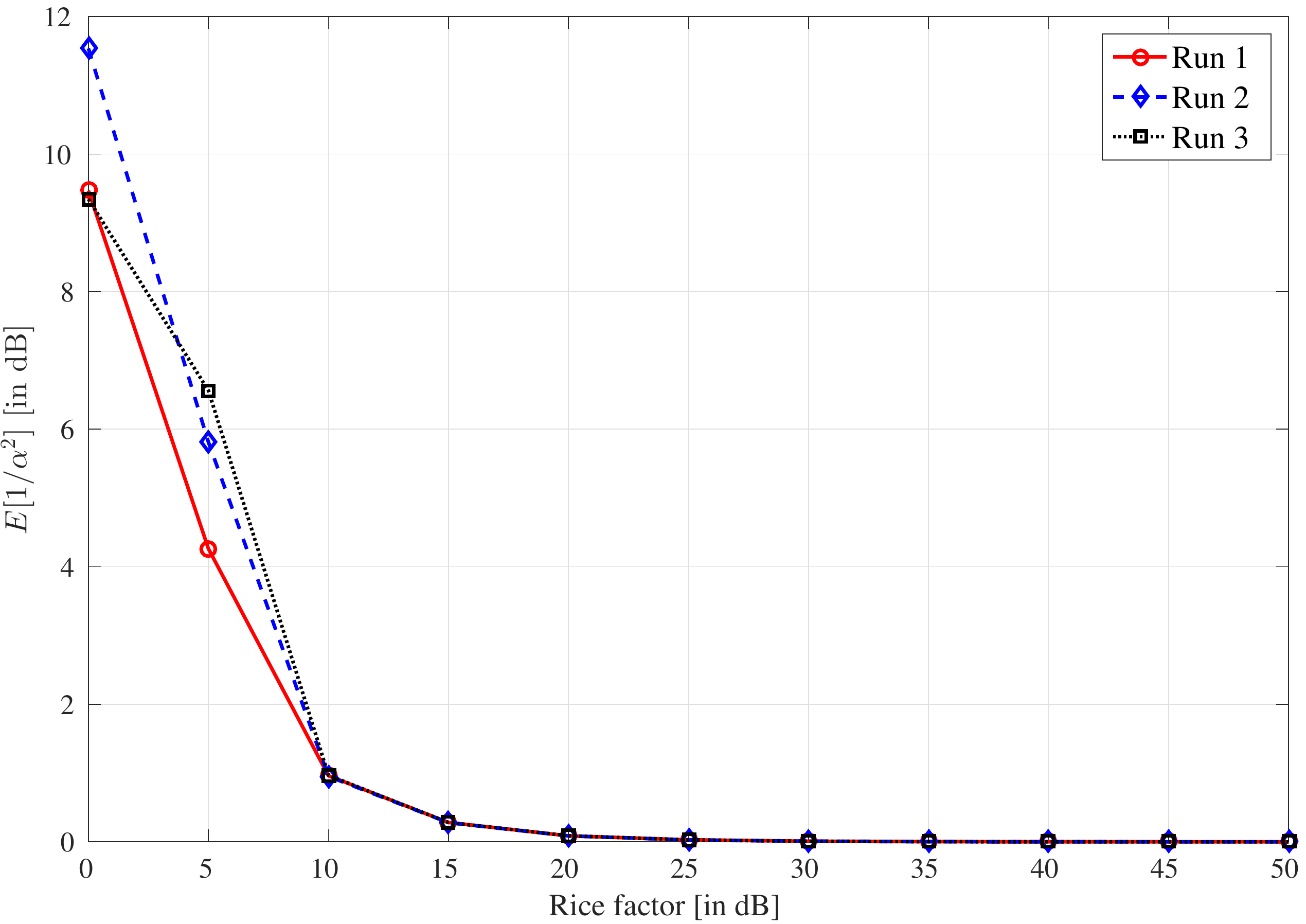}
\caption{The figure shows the inverse second moment of a Rician random variable versus Rice factor for three independent runs.}
%. The Rice factor $K$ was set equal to $10$ dB. In all the cases ratio of number of pilots to number of parameters estimated was kept equal to $50$.}
\label{fig_inv_sec_mom}
\end{center}
\end{figure}

Following remarks are in order with respect to Proposition \ref{prop:b_crlb_sp_case}: (i) For our numerical results evaluation of $L$ non-zero terms of the matrix $\mathbf{J}_{\text{P}}$ is done using Monte-Carlo (MC) simulations. (ii) For $K_{l}=0$ (i.e. the Rayleigh fading case) the CRLB becomes undefined, since in this case the non-zero terms of matrix $\mathbf{J}_{\text{P}}$ will be equal to $\mathbb{E}[\frac{1}{\alpha_{l}^{2}}]=\int_{0}^{\infty} \frac{1}{\alpha_{l}^{2}} \frac{\alpha_{l}}{\sigma_{l}^{2}} e^{-\frac{\alpha_{l}^{2}}{\sigma_{l}^{2}}} d\alpha_{l}$, which diverges to $\infty$ (since inverting a Rayleigh random variable would require infinite amount of power). This behavior can also be inferred from Fig. \ref{fig_inv_sec_mom}, in which we plot the magnitude of the inverse second moment of a Rician random variable for different values of Rice factor for three independent runs. For every run the average is computed via MC simulation by drawing $100000$ randomly generated realizations. As can be seen from the Fig. \ref{fig_inv_sec_mom} for small values of Rice factor the magnitude of the inverse second moment is relatively large compared to the case when the Rice factor is large.  Further the value of the inverse second moment for smaller values of Rice factor are unstable, as can be inferred from the variation in the corresponding value for three different runs. However, for larger values of Rice factor the value is much more stable and indeed as the Rice factor gets larger it converges to $1$.

\begin{figure}[p]
\begin{center}
\includegraphics[width=0.95\linewidth]{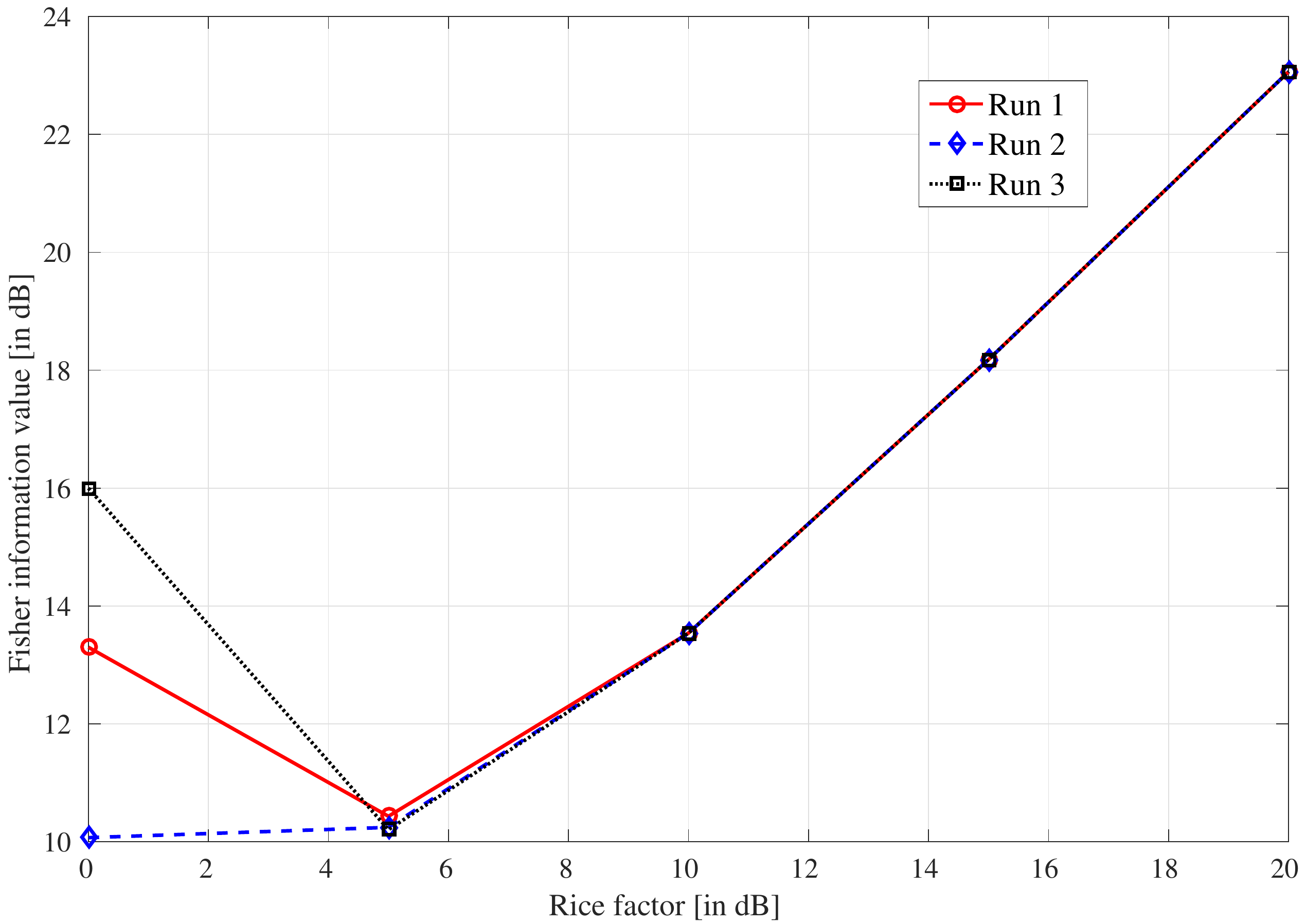}
\caption{The figure shows the Fisher information value of a Rician random variable versus Rice factor for three independent runs.}
%. The Rice factor $K$ was set equal to $10$ dB. In all the cases ratio of number of pilots to number of parameters estimated was kept equal to $50$.}
\label{fig_fim_rice_rv}
\end{center}
\end{figure}

The instability or oscillatory behavior observed in Fig. \ref{fig_inv_sec_mom} can also be observed in Fig. \ref{fig_fim_rice_rv}, in which we plot the Fisher information value (i.e. $\mathbb{E}[-\frac{\partial^{2}\ln(p(\alpha_{l}))}{\partial \alpha_{l}^{2}}]$, $\mathbb{E}[.]$ w.r.t $p(\alpha_{l})$ - Rician PDF, whose expression is given by \cite[Eq. 2.3-61]{PS2008}) of a Rician random variable computed via MC simulations by drawing $100000$ randomly generated realizations for three independent runs.
\section{Numerical Results}
\label{sect:numresults}
In this section we plot the derived Bayesian CRLB versus the signal-to-noise ratio (SNR) for different parameter values of interest.
\subsection{Description of Parameter Values}
\label{subsect:numresults_description}
The SNR at the receiver is equal to $\frac{n_t n_r \sum_{l=1}^{L}\Omega_{l}}{\sigma_{v}^{2}}$ for a normalized transmit power of unity, where $\Omega_{l} \triangleq \mathbb{E}(\alpha_{l}^{2})$. For our numerical results we normalize the sum of average power gains across all paths to be equal to unity i.e. $\Omega \triangleq \sum_{l=1}^{L}\Omega_{l}^{2}=1$, therefore in our case $\text{SNR}=\frac{n_t n_r}{\sigma_{v}^{2}}$. Since we assumed $\alpha_l~(l=1,\cdots,L)$ to be Ricean distributed therefore let $K_{l} \triangleq \frac{\mu_{l}^{2}}{\sigma_{l}^{2}}$ denote the Rice factor of path $l$. For an $\alpha_l$ of Ricean distribution $\Omega_{l}=\mu_{l}^{2} + \sigma_{l}^{2}=\sigma_{l}^{2}\left(1 + K_{l}\right)$ \cite{PS2008}. If we assume an exponential power delay profile for the multi-path channel, then $\Omega_{l}=\Omega_{1}e^{-(l-1)\delta}$, where $\delta$ is the decay parameter associated with the power delay profile. Since we normalized $\Omega$ to unity, therefore we obtain $\Omega_{1}=\frac{1}{\sum_{l=1}^{L} e^{-(l-1)\delta}}$. Thus for a given value of $L$ and $\delta$ one can determine $\Omega_1$ and thereby set $\sigma_{l}^{2}=\frac{\Omega_{1} e^{-(l-1)\delta}}{(1 + K_{l})}$. For the purpose of our numerical results we set $n_t=16$, $n_r=16$ and $\delta=0.5$.

The beamforming and beamcombining matrices ($\mathbf{F}$ and $\mathbf{G}$) are generated in three different methods as described in the following. The first one referred to as {\em{non-uniform}} method divides the range of $\phi_{l}$ and $\psi_{l}$ (for every $l=1,\cdots,L$) respectively i.e. $[-1,1]$, into $p_t$ and $p_r$ bins respectively. The quantized beamforming/beamcombining direction is chosen to be equal to the arccosine of the center of each bin. This method is tantamount to non-uniform quantization and hence the name. The second method referred to as {\em{uniform}} divides the range between $0$ and $\pi$ into $p_t$ and $p_r$ bins. The quantized beamforming/beamcombining direction is chosen to be equal to the center of each bin. Finally, in the third method referred to as {\em{orthogonal}} we choose the respective columns of $\mathbf{F}$ and $\mathbf{G}$ so that they are orthogonal to each other, thereby implying $\mathbf{K}=\mathbf{I}_{p_{t} p_{r}}$.

Note that for Fig. \ref{fig_crlb_snr_diff_ricefactor_values} and Fig. \ref{fig_crlb_ricefactor_diff_snr} we assume the beam forming and combining matrices to be generated according to the non-uniform method with $p_{t}=p_{r}=n_{t}=n_{r}=16$. For Fig. \ref{fig_crlb_pilots_diff_beamformers} we set the SNR equal to $15$ dB. For the purpose of our numerical results study in Fig. \ref{fig_crlb_ricefactor_diff_snr} we introduce the following parameter: PPR $\triangleq$ No. of pilot symbols/No. of estimated parameters.
\subsection{Discussion on numerical results}
\begin{figure}[p]
\begin{center}
\includegraphics[width=0.95\linewidth]{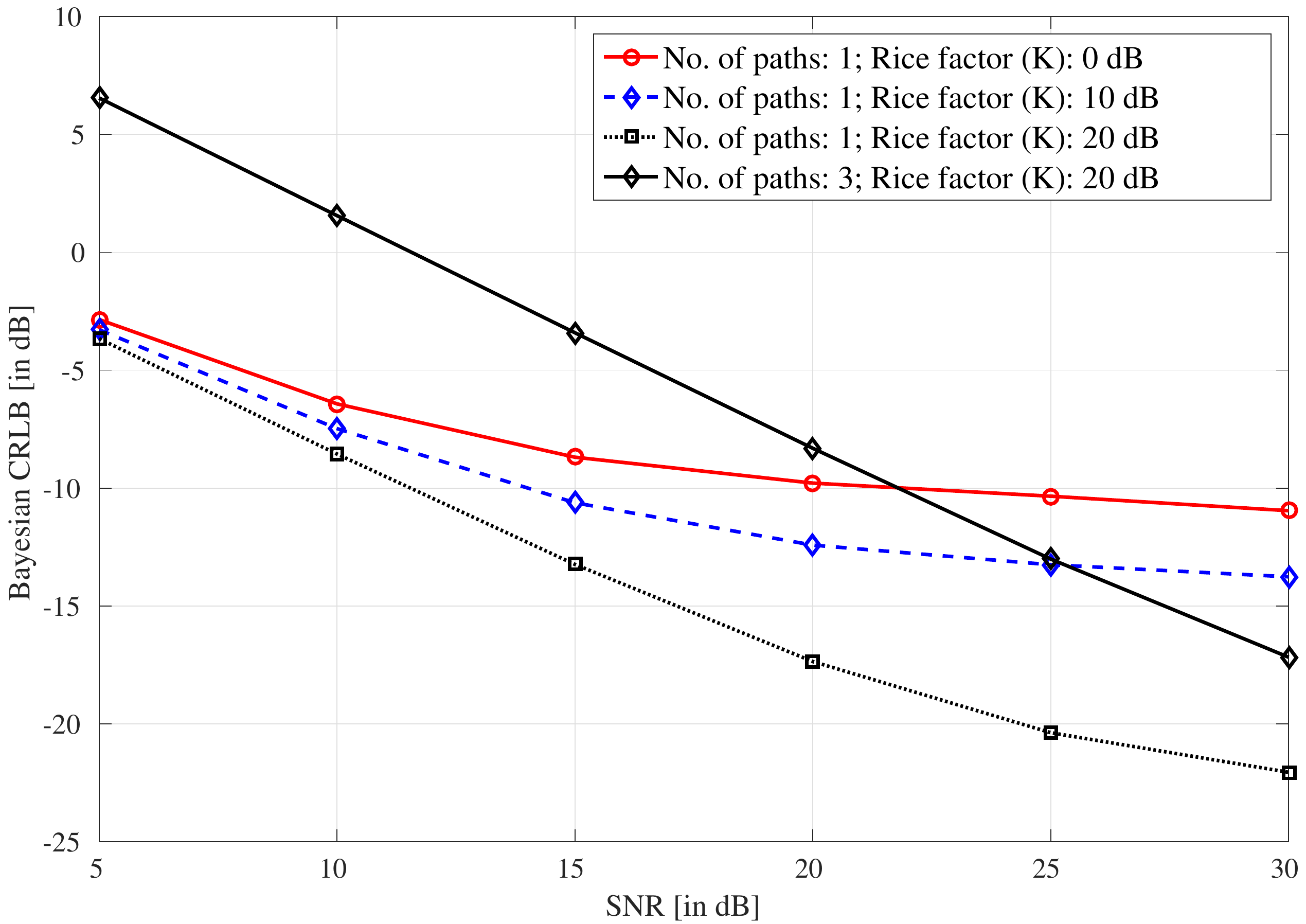}
\caption{The figure shows the derived Bayesian CRLB versus SNR for different values of the Rice factor.}
\label{fig_crlb_snr_diff_ricefactor_values}
\end{center}
\end{figure}

\begin{figure}[p]
\begin{center}
\includegraphics[width=0.95\linewidth]{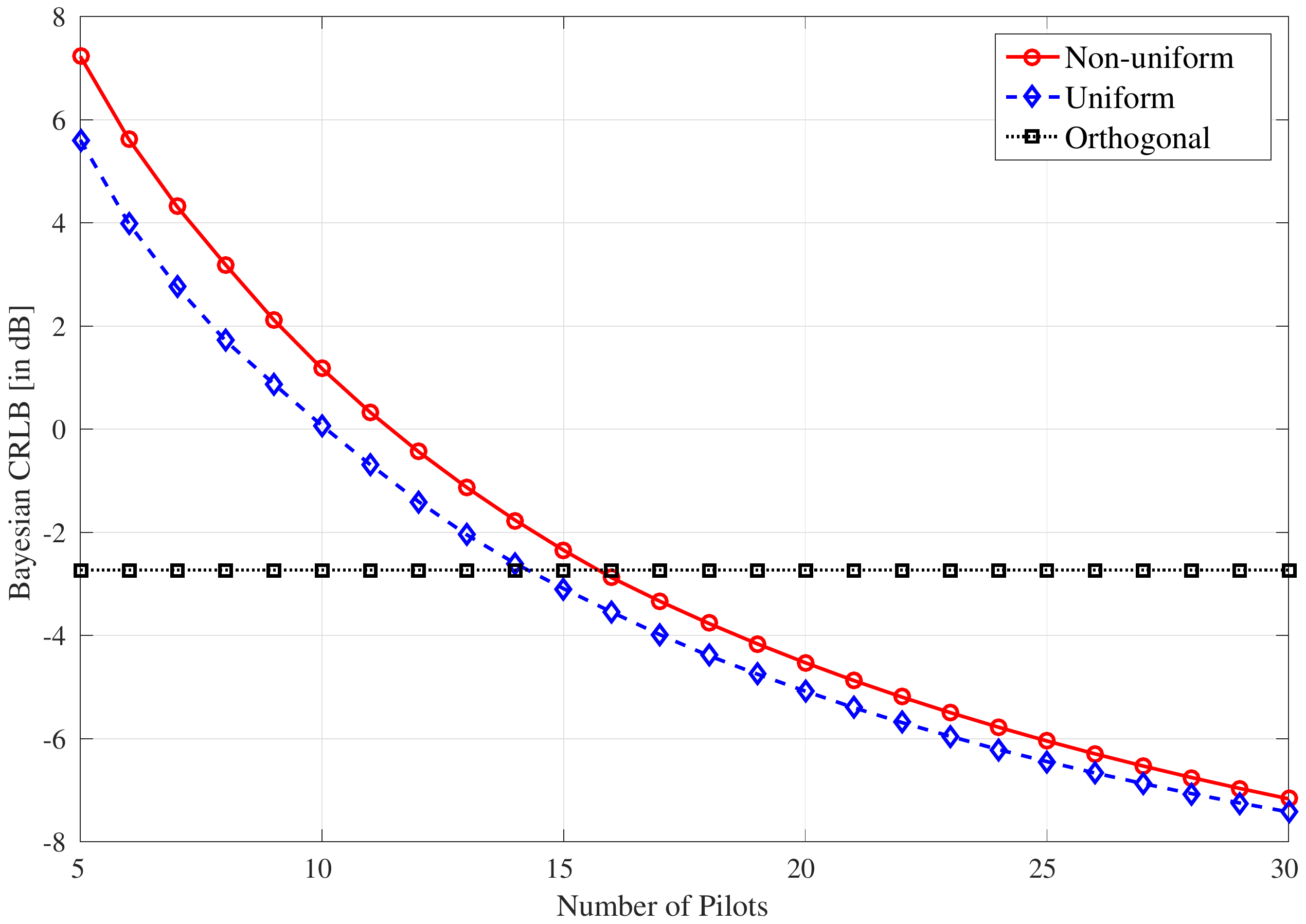}
\caption{The figure shows the derived Bayesian CRLB versus number of pilots for different types of beamforming code books.} %The number of paths were set equal to $3$, the Rice factor was set equal to $0$ dB and SNR was set equal to $15$ dB.}
\label{fig_crlb_pilots_diff_beamformers}
\end{center}
\end{figure}

% \begin{figure}[t]
% \begin{center}
% \includegraphics[width=0.95\linewidth]{crlb_vs_snr_for_different_paths.eps}
% %\epsfig{figure=algo1-1.eps, width=1.50\linewidth}
% %\include{figure=algo1.eps, width=1.00\linewidth}
% \caption{The figure shows the derived Bayesian CRLB versus SNR for different number of paths. The Rice factor was set equal to $0$ dB.}
% \label{fig_crlb_snr_diff_paths}
% \end{center}
% \end{figure}
Fig. \ref{fig_crlb_snr_diff_ricefactor_values} contains plots of Bayesian CRLB expression versus SNR for different values of Rice factor. As can be seen from the plots, in general as SNR increases, CRLB decreases. Also the CRLB for Rice factor equal to $20$ dB is strictly lower than the corresponding CRLB values for Rice factor values equal to $10$ dB and $0$ dB respectively. The reason for this trend can be explained as follows: an increase in Rice factor moves the channel behavior towards that of an AWGN channel, whose CRLB is lower than that of any fading channel. Further for fixed values of Rice factor and number of pilots, the CRLB for $L=3$ is strictly above that of the case when $L=1$. This is because for a fixed pilot budget, the number of parameters to be estimated increase with an increase in number of paths.

Fig. 2 contains plots of Bayesian CRLB versus the number of pilots for different beamformer types. As can be seen the CRLB of the uniform beamformer is strictly below that of the non-uniform beamformer across the entire range of values for number of pilots, although for smaller values of number of pilots the orthogonal beamformer performs better than the other two. This gives important insights into how the beam forming and combining code books be designed. Further for all the three types of beamformers, as expected, the CRLB decreases gradually with an increase in number of pilots.

\begin{figure}[t]
\begin{center}
\includegraphics[width=0.95\linewidth]{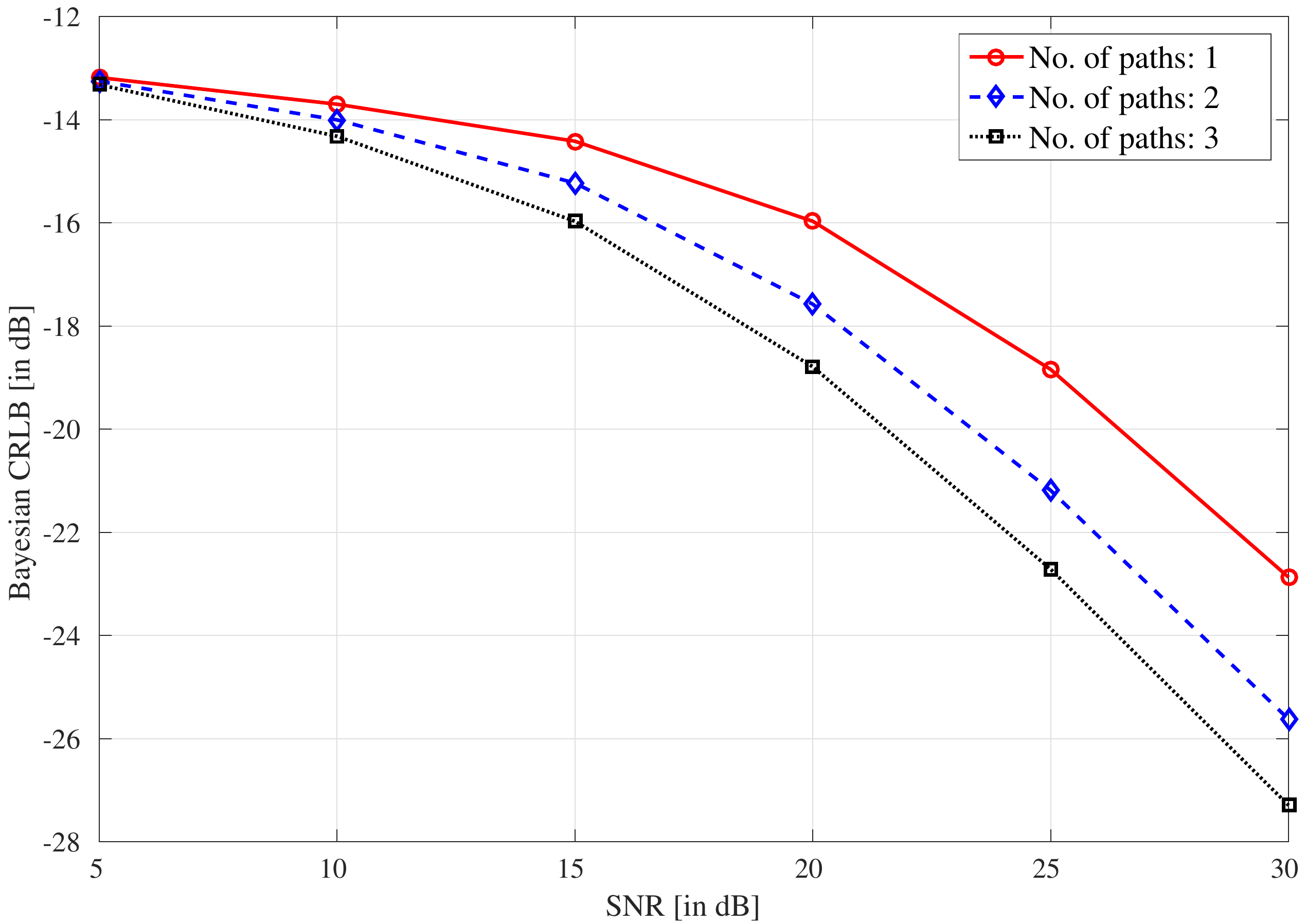}
\caption{The figure shows the derived Bayesian CRLB versus SNR for different values of number of paths and fixed PPR$=50$.}
%. The Rice factor $K$ was set equal to $10$ dB. In all the cases ratio of number of pilots to number of parameters estimated was kept equal to $50$.}
\label{fig_crlb_ricefactor_diff_snr}
\end{center}
\end{figure}

Finally, Fig. \ref{fig_crlb_ricefactor_diff_snr} contains plots of Bayesian CRLB versus SNR for a fixed value of PPR$=50$. As can be seen from the plots as the number of paths $L$ increase from $1$ to $3$ CRLB decreases across all the SNR values. This is in contrast to what was observed in Fig. \ref{fig_crlb_snr_diff_ricefactor_values} this somewhat of a contradictory behavior can be explained as follows. For the same value of PPR as the number of paths increase, the number of estimated parameters increase, which thereby cause the number of pilots used to also increase, thus translating into better performance in terms of achieving a lower CRLB.

% \begin{figure}[t]
% \begin{center}
% \includegraphics[width=0.95\linewidth]{crlb_vs_snr_for_different_beamformers.eps}
% %\epsfig{figure=algo1-1.eps, width=1.50\linewidth}
% %\include{figure=algo1.eps, width=1.00\linewidth}
% \caption{The figure shows the derived Bayesian CRLB versus SNR for different types of beamforming code books. The number of paths were set equal to $3$ and the Rice factor was set equal to $0$ dB.}
% \label{fig_crlb_snr_diff_beamformers}
% \end{center}
% \end{figure}
% \begin{figure}[t]
% \begin{center}
% \includegraphics[width=0.95\linewidth]{crlb_vs_snr_for_different_paths_fixed_pilotratio.pdf}
% %\epsfig{figure=algo1-1.eps, width=1.50\linewidth}
% %\include{figure=algo1.eps, width=1.00\linewidth}
% \caption{The figure shows the derived Bayesian CRLB versus SNR for different values of number of paths. The Rice factor $K$ was set equal to $10$ dB. In all the cases ratio of number of pilots to number of parameters estimated was kept equal to $50$.}
% \label{fig_crlb_ricefactor_diff_snr}
% \end{center}
% \end{figure}
\section{Conclusions}
\label{sect:conclusions}
In this report we derived the random and Bayesian CRLB for the joint estimation problem of AoA, AoD and multi-path gains. The proposed CRLB shall be useful for comparing the performance of various practical estimators. Our numerical results based on the derived CRLB point to the following important observations: (i) an increase in Rice factor in general decreases the CRLB and (ii) the combination of beam forming and beam combining matrices generated by quantizing the domain of AoA and AoD (=$[0,\pi)$) directly yields a lower CRLB than other methods. As part of our future work we shall consider extending the joint estimation problem considered in this report by including the elevation angle at both transmitter and receiver.
\newpage
\section*{Appendix A: Calculation of various entries of $\mathbf{J}_{\text{NR}}$}
% \section{Calculation of Fisher Information Matrix (FIM)}
% Given a parameter $\boldsymbol{\theta} \triangleq \left[\theta_{1},\cdots,\theta_{m}\right]^{T}$ the $(i,j)^{\text{th}}$ entry of FIM $I_{\boldsymbol{\theta}}$ is given by 
% \begin{equation}
% I_{\boldsymbol{\theta}}(i,j)=-\mathbb{E}\left[\frac{\partial^{2} L(\mathbf{y})}{\partial \theta_{i} \partial \theta_{j}}\right]\label{fim_def_eqn}
% \end{equation}
The following properties of ``$\text{tr}[.]$'' and Kronecker product ``$\otimes$'' operators respectively \cite[Appendix A]{vantrees_part4} would be useful in the ensuing calculations related to the various entries of $\mathbf{J}_{\text{NR}}$.
\begin{eqnarray}
\mathbf{x}^{\dagger} \mathbf{\Sigma} \mathbf{y}&=&\text{tr}\left[\mathbf{\Sigma} \mathbf{y} \mathbf{x}^{\dagger}\right]\label{trace_op_property}\\
\left(\mathbf{U} \otimes \mathbf{V})(\mathbf{X} \otimes \mathbf{Y}\right)&=&(\mathbf{U}\mathbf{X}) \otimes (\mathbf{V}\mathbf{Y})\label{kron_prod_property}
\end{eqnarray}

\subsection*{Calculation of: $\mathbf{J}_{\text{NR}}\left(\phi_{l},\phi_{l}\right) \triangleq  -\mathbb{E}\left[\frac{\partial^{2} L(\mathbf{y})}{\partial \phi_{l}^{2}}\right]$}
Note that he log-likelihood function $L(\mathbf{y})$ of the observation model in Eq. (\ref{observation_vector_def}) is given by Eq. (\ref{llr_eqn}).
Differentiating $L(\mathbf{y})$ with respect to (w.r.t) $\phi_{l}$ we obtain
\begin{equation}
%\frac{\partial L(\mathbf{y})}{\partial \phi_{l}}&=&\frac{1}{\sigma_{v}^{2}}\frac{\partial}{\partial \phi_{l}} \left[\mathbf{y}^{H}\mathbf{m} + \mathbf{m}^{H}\mathbf{y} - \mathbf{m}^{H}\mathbf{m}\right]\nonumber\\
\frac{\partial L(\mathbf{y})}{\partial \phi_{l}}=\frac{1}{\sigma_{v}^{2}}\left[\left(\mathbf{y} - \mathbf{m}\right)^{\dagger}\frac{\partial \mathbf{m}}{\partial \phi_{l}} + \frac{\partial \mathbf{m}^{\dagger}}{\partial \phi_{l}}\left(\mathbf{y} - \mathbf{m}\right)\right]\label{eqn_fim_phi_1}
\end{equation}
Differentiating Eq. (\ref{eqn_fim_phi_1}) again w.r.t $\phi_{l}$ we obtain
\begin{eqnarray}
&&\frac{\partial^{2} L(\mathbf{y})}{\partial \phi_{l}^{2}}=\nonumber\\
&&\frac{1}{\sigma_{v}^{2}}\left[-\frac{\partial \mathbf{m}^{\dagger}}{\partial \phi_{l}}\frac{\partial \mathbf{m}}{\partial \phi_{l}} + \left(\mathbf{y} - \mathbf{m}\right)^{\dagger}\frac{\partial^{2} \mathbf{m}}{\partial \phi_{l}^{2}} + \frac{\partial^{2} \mathbf{m}^{\dagger}}{\partial \phi_{l}^{2}}\left(\mathbf{y} - \mathbf{m}\right) - \frac{\partial \mathbf{m}^{\dagger}}{\partial \phi_{l}}\frac{\partial \mathbf{m}}{\partial \phi_{l}}~. \right]\label{eqn_fim_phi_2}
\end{eqnarray}
Taking $\mathbb{E}$ of Eq. (\ref{eqn_fim_phi_2}) w.r.t $p\left(\mathbf{y}|\boldsymbol{\theta}\right)$, we obtain
\begin{equation}
\mathbf{J}_{\text{NR}}\left(\phi_{l},\phi_{l}\right)=\frac{2}{\sigma_{v}^{2}}\frac{\partial \mathbf{m}^{\dagger}}{\partial \phi_{l}}\frac{\partial \mathbf{m}}{\partial \phi_{l}}~.\label{eqn_fim_phi_3}
\end{equation}
Next we shall compute $\frac{\partial \mathbf{m}^{\dagger}}{\partial \phi_{l}}$. From the definition of $\mathbf{m}$ in Eq. (\ref{observation_vector_def}) we have
\begin{eqnarray}
&&\mathbf{m}^{\dagger}=\sum_{l=1}^{L}\alpha_{l}\left(\mathbf{e}_{t}^{T}(\phi_l) \otimes \mathbf{e}_{r}^{\dagger}(\psi_l)\right) \mathbf{A}^{\dagger}\nonumber\\
&&\frac{\partial \mathbf{m}^{\dagger}}{\partial \phi_{l}}=\alpha_{l}\frac{\partial}{\partial \phi_{l}}\left(\mathbf{e}_{t}^{T}(\phi_l) \otimes \mathbf{e}_{r}^{\dagger}(\psi_l)\right) \mathbf{A}^{\dagger}\nonumber\\
&&\vdots\nonumber\\
&&=\alpha_{l} j \pi \sin(\phi_{l})\left(\widetilde{\mathbf{e}}_{t}^{\dagger}(\phi_l) \otimes \mathbf{e}_{r}^{\dagger}(\psi_l)\right)\mathbf{A}^{\dagger}~,\label{eqn_fim_phi_4}
\end{eqnarray}
where $\widetilde{\mathbf{e}}_{t}$ is defined as in Section \ref{subsect:crlb_llr_def}.
%$\tilde{\mathbf{e}}_{t} \triangleq \frac{1}{\sqrt{n_t}}\left[0,e^{j \pi \cos(\phi_{l})},2e^{j 2 \pi \cos(\phi_{l})},\cdots,(n_{t}-1)e^{j (n_{t} -1) \pi \cos(\phi_{l})}\right]^{T}$.

Similarly we can show
\begin{equation}
\frac{\partial \mathbf{m}}{\partial \phi_{l}}=-A\alpha_{l} j \pi \sin(\phi_{l})\left(\widetilde{\mathbf{e}}_{t} (\phi_{l}) \otimes \mathbf{e}_{r}(\psi_{l})\right)\label{eqn_fim_phi_5}
\end{equation}
Using Eq. (\ref{eqn_fim_phi_4}) and Eq. (\ref{eqn_fim_phi_5}) in Eq. (\ref{eqn_fim_phi_3}) and noting that $\mathbf{K} \triangleq \mathbf{A}^{\dagger}\mathbf{A}$, we obtain
\begin{eqnarray}
\mathbf{J}_{\text{NR}}\left(\phi_{l},\phi_{l}\right)&=&\frac{2 \alpha_{l}^{2}}{\sigma_{v}^{2}} \pi^{2} \sin^{2}(\phi_l) (\widetilde{\mathbf{e}}_{t}^{\dagger}(\phi_l) \otimes \mathbf{e}_{r}^{\dagger}(\psi_l)) \mathbf{K} (\widetilde{\mathbf{e}}_{t} (\phi_l) \otimes \mathbf{e}_{r}(\psi_l))~, \nonumber\\
&=&\frac{2 \alpha_{l}^{2}}{\sigma_{v}^{2}} \pi^{2} \sin^{2}(\phi_l)\text{tr}\left[\mathbf{K}(\widetilde{\mathbf{e}}_{t} (\phi_l) \otimes \mathbf{e}_{r}(\psi_l))(\widetilde{\mathbf{e}}_{t}^{\dagger}(\phi_l) \otimes \mathbf{e}_{r}^{\dagger}(\psi_l))\right]\label{eqn_fim_phi_6}
\end{eqnarray}
Using the property of $\text{tr}[.]$ and $\otimes$ given in Eq. (\ref{trace_op_property}) and Eq. (\ref{kron_prod_property}), we can simplify Eq. (\ref{eqn_fim_phi_6}) as follows:
\begin{eqnarray}
\mathbf{J}_{\text{NR}}\left(\phi_{l},\phi_{l}\right)&=&\frac{2 \pi^2}{\sigma_{v}^{2}}\alpha_{l}^{2}\text{tr}\left(\mathbf{K} \left(\mathbf{P} \otimes \mathbf{Q}\right) \right)\label{eqn_fim_phi_7}\\
\text{where}~ \mathbf{P} &\triangleq& \sin^{2}(\phi_{l})\widetilde{\mathbf{e}}_{t}(\phi_{l}) \widetilde{\mathbf{e}}^{\dagger}_{t}(\phi_{l})~ \text{and}~ \mathbf{Q} \triangleq \mathbf{e}_{r}(\psi_{l}) \mathbf{e}^{\dagger}_{r}(\psi_{l})\label{pq_matrix_def}.
\end{eqnarray}
Using the definition of $\widetilde{\mathbf{e}}_{t}(.)$ and $\mathbf{e}_{r}(.)$, the general element in matrices $\mathbf{P}$ and $\mathbf{Q}$ can be written as follows:
\begin{eqnarray}
\mathbf{P}(r,s)&=&\frac{(r-1)(s-1)}{n_{t}}e^{-j\pi(s-r)\cos(\phi_{l})} \sin^{2}(\phi_{l})~\text{and}\label{p_matrix_gen_element_def}\\
%&&Q_{m,n}=\frac{1}{n_{r}}J_{0}\left(\pi(n-m)\right)\label{bayesian_fim_phi_eqn_5}
\mathbf{Q}(r,s)&=&\frac{1}{n_{r}}e^{j \pi(s-r)\cos(\psi_{l})}\label{q_matrix_gen_element_def}
\end{eqnarray}
%where $\tilde{\mathbf{e}}_{t}$ is defined as under Eq. (\ref{eqn_fim_phi_4}).
\subsection*{Calculation of: $\mathbf{J}_{\text{NR}}\left(\phi_{l},\alpha_{l}\right) \triangleq -\mathbb{E}\left[\frac{\partial^{2} L(\mathbf{y})}{\partial \phi_{l} \partial \alpha_{l}}\right]$}
Differentiating Eq. (\ref{eqn_fim_phi_1}) w.r.t $\alpha_{l}$ and subsequently applying $-\mathbb{E}$ w.r.t $p\left(\mathbf{y}|(.),(.),(.)\right)$, we obtain
\begin{equation}
\mathbf{J}_{\text{NR}}\left(\phi_{l},\alpha_{l}\right) =\frac{1}{\sigma_{v}^{2}}\left[\frac{\partial \mathbf{m}^{\dagger}}{\partial \phi_{l}}\frac{\partial \mathbf{m}}{\partial \alpha_{l}} + \frac{\partial \mathbf{m}^{\dagger}}{\partial \alpha_{l}}\frac{\partial \mathbf{m}}{\partial \phi_{l}}\right]\label{eqn_fim_phi_alpha_1}
\end{equation}
Differentiating $\mathbf{m}$ defined in Eq. (\ref{observation_vector_def}) w.r.t $\alpha_{l}$ we obtain
\begin{equation}
\frac{\partial \mathbf{m}}{\partial \alpha_{l}}=\mathbf{A} \left(\mathbf{e}_{t}^{*}(\phi_{l}) \otimes \mathbf{e}_{r}(\psi_{l})\right) \Rightarrow \frac{\partial \mathbf{m}^{\dagger}}{\partial \alpha_{l}}= \left(\mathbf{e}_{t}^{T}(\phi_{l}) \otimes \mathbf{e}_{r}^{\dagger}(\psi_{l})\right) \mathbf{A}^{\dagger}\label{eqn_fim_phi_alpha_2}
\end{equation}
Using Eq. (\ref{eqn_fim_phi_4}), Eq. (\ref{eqn_fim_phi_5}) and Eq. (\ref{eqn_fim_phi_alpha_2}) in Eq. (\ref{eqn_fim_phi_alpha_1}) and further noting that $\mathbf{K} \triangleq \mathbf{A}^{\dagger}\mathbf{A}$, we obtain
\begin{eqnarray}
\mathbf{J}_{\text{NR}}\left(\phi_{l},\alpha_{l}\right)&=&\frac{-2 \alpha_{l} \pi \sin(\phi_{l})}{\sigma_{v}^{2}} \text{Im}\left[\left(\widetilde{\mathbf{e}}_{t}^{\dagger}(\phi_l) \otimes \mathbf{e}_{r}^{\dagger}(\psi_l)\right)\mathbf{A}^{\dagger} \mathbf{A} \left(\mathbf{e}_{t}^{*}(\phi_{l}) \otimes \mathbf{e}_{r}(\psi_{l})\right)\right]~,\nonumber\\
&=&\frac{-2 \alpha_{l} \pi \sin(\phi_{l})}{\sigma_{v}^{2}} \text{Im}\left[\left(\widetilde{\mathbf{e}}_{t}^{\dagger}(\phi_l) \otimes \mathbf{e}_{r}^{\dagger}(\psi_l)\right)\mathbf{K} \left(\mathbf{e}_{t}^{*}(\phi_{l}) \otimes \mathbf{e}_{r}(\psi_{l})\right)\right].\label{eqn_fim_phi_alpha_3}
\end{eqnarray}
where $\text{Im}[z]$ denotes the imaginary part of the complex number $z$.

Using the property of $\text{tr}[.]$ and $\otimes$ given in Eq. (\ref{trace_op_property}) and Eq. (\ref{kron_prod_property}) we can simplify Eq. (\ref{eqn_fim_phi_alpha_3}) as follows:
\begin{eqnarray}
\mathbf{J}_{\text{NR}}\left(\phi_{l},\alpha_{l}\right)&=&\frac{-2 \pi \alpha_{l}}{\sigma_{v}^{2}} \text{Im}\left\{\text{tr}\left[\mathbf{K} \left(P^{(2)} \otimes Q\right) \right]\right\}\\
\text{where}~ \mathbf{P}^{(2)} &\triangleq& \sin(\phi_{l})\mathbf{e}^{*}_{t}(\phi_{l}) \widetilde{\mathbf{e}}^{\dagger}_{t}(\phi_{l})\label{p2_matrix_def}
\end{eqnarray}
and $\mathbf{Q}$ is defined as in Eq. (\ref{pq_matrix_def}).

The general element of matrix $\mathbf{P}^{(2)}$ can be written as follows:
\begin{equation}
\mathbf{P}^{(2)}(r,s)=\frac{(s-1)}{n_{t}}e^{-j\pi(s-r)\cos(\phi_{l})} \sin(\phi_{l})\label{p2_general_element_def}.
\end{equation}
\subsection*{Calculation of: $\mathbf{J}_{\text{NR}}\left(\phi_{l},\psi_{l}\right) \triangleq -\mathbb{E}\left[\frac{\partial^{2} L(\mathbf{y})}{\partial \phi_{l} \partial \psi_{l}}\right]$}
Differentiating Eq. (\ref{eqn_fim_phi_1}) w.r.t $\psi_{l}$ and subsequently applying $-\mathbb{E}$ w.r.t $p\left(\mathbf{y}|(.),(.),(.)\right)$, we obtain
\begin{equation}
\mathbf{J}_{\text{NR}}\left(\phi_{l},\psi_{l}\right)=\frac{1}{\sigma_{v}^{2}} \left[\frac{\partial \mathbf{m}^{\dagger}}{\partial \phi_{l}}\frac{\partial \mathbf{m}}{\partial \psi_{l}} + \frac{\partial \mathbf{m}^{\dagger}}{\partial \psi_{l}}\frac{\partial \mathbf{m}}{\partial \phi_{l}}\right]\label{eqn_fim_phi_psi_1}
\end{equation}
The quantity
\begin{equation}
\frac{\partial \mathbf{m}}{\partial \psi_{l}}=\mathbf{A} \alpha_{l} j \pi \sin(\psi_{l}) \left(\mathbf{e}_{t}^{*}(\phi_{l}) \otimes \widetilde{\mathbf{e}}^{*}_{r}(\psi_l)\right)~,\label{eqn_fim_phi_psi_2}
\end{equation}
where $\widetilde{\mathbf{e}}_{r}(\psi_l) = \frac{1}{\sqrt{n_r}}\left[0,e^{j \pi \cos(\psi_{l})},2e^{j 2\pi \cos(\psi_{l})},\cdots,(n_{r}-1)e^{j (n_{r}-1)\pi \cos(\psi_{l})}\right]^{T}$.

The Eq. (\ref{eqn_fim_phi_psi_2}) implies that
\begin{equation}
\frac{\partial \mathbf{m}^{\dagger}}{\partial \psi_{l}}=-j \alpha_{l}  \pi \sin(\psi_{l})\left(\mathbf{e}_{t}^{T}(\phi_{l}) \otimes \widetilde{\mathbf{e}}^{T}_{r}(\psi_l)\right)\mathbf{A}^{\dagger}  ~,\label{eqn_fim_phi_psi_3}
\end{equation}
Using Eq. (\ref{eqn_fim_phi_4}), Eq. (\ref{eqn_fim_phi_5}), Eq. (\ref{eqn_fim_phi_psi_2}) and Eq. (\ref{eqn_fim_phi_psi_3}) in Eq. (\ref{eqn_fim_phi_psi_1}), and noting that $\mathbf{K} \triangleq \mathbf{A}^{\dagger}\mathbf{A}$, we obtain
\begin{eqnarray}
&&\mathbf{J}_{\text{NR}}\left(\phi_{l},\psi_{l}\right)=\frac{-\pi^{2}\alpha_{l}^{2}\sin(\psi_{l})\sin(\phi_{l})}{\sigma_{v}^{2}}\left[\text{term 1}~ + \text{term 2}\right]\nonumber\\
&&\text{where}~\nonumber\\
&&\text{term 1}~=\left(\mathbf{e}_{t}^{T}(\phi_{l}) \otimes \widetilde{\mathbf{e}}^{T}_{r}(\psi_l)\right)\mathbf{K}\left(\widetilde{\mathbf{e}}_{t}(\phi_{l}) \otimes \mathbf{e}_{r}(\psi_l)\right),\nonumber\\
&&\text{term 2}~=\left(\widetilde{\mathbf{e}}_{t}^{\dagger}(\phi_{l}) \otimes \mathbf{e}^{\dagger}_{r}(\psi_l)\right) \mathbf{K} \left(\mathbf{e}_{t}^{*}(\phi_{l}) \otimes \widetilde{\mathbf{e}}^{*}_{r}(\psi_l)\right)\nonumber\\
&&=\frac{-2 \pi^{2}\alpha_{l}^{2}\sin(\psi_{l})\sin(\phi_{l})}{\sigma_{v}^{2}}~\text{Re}\left[\left(\mathbf{e}_{t}^{T}(\phi_{l}) \otimes \widetilde{\mathbf{e}}^{T}_{r}(\psi_l)\right)\mathbf{K}\left(\widetilde{\mathbf{e}}_{t}(\phi_{l}) \otimes \mathbf{e}_{r}(\psi_l)\right)\right]~,\label{eqn_fim_phi_psi_4}
\end{eqnarray}
where ``$\text{Re}$'' denotes the real part of a complex number.

By using the property of $\text{tr}[.]$ and $\otimes$ given in Eq. (\ref{trace_op_property}) and Eq. (\ref{kron_prod_property}), it is straightforward to show that
\begin{eqnarray}
\mathbf{J}_{\text{NR}}\left(\phi_{l},\psi_{l}\right) &=& \frac{-2 \pi^{2}\alpha_{l}^{2}}{\sigma_{v}^{2}}~\text{Re}\left\{\text{tr}\left[\mathbf{K} \left(\mathbf{P}^{(3)}\otimes \mathbf{Q}^{(2)}\right)\right]\right\},\label{eqn_fim_phi_psi_5}\\
\text{where}~\mathbf{P}^{(3)} &\triangleq& \sin(\phi_{l})\widetilde{\mathbf{e}}_{t}(\phi_{l})\mathbf{e}_{t}^{T}(\phi_{l})\label{p3_matrix_def}\\
\text{and}~\mathbf{Q}^{(2)} &\triangleq& \sin(\psi_{l})\mathbf{e}_{r}(\psi_{l})\widetilde{\mathbf{e}}_{r}^{T}(\psi_{l})~.\label{q2_matrix_def}
\end{eqnarray}
The general element in matrix $\mathbf{P}^{(3)}$ and $\mathbf{Q}^{(2)}$ can be written as follows:
\begin{eqnarray}
\mathbf{P}^{(3)}(r,s)&=&\frac{(r-1)}{n_{t}}e^{-j\pi(s-r)\cos(\phi_{l})} \sin(\phi_{l})\label{p3_general_element_def}\\
\mathbf{Q}^{(2)}(r,s)&=&\frac{(s-1)}{n_{r}}e^{-j\pi(r-s)\cos(\psi_{l})} \sin(\psi_{l})\label{q2_general_element_def}
\end{eqnarray}
\subsection*{Calculation of: $\mathbf{J}_{\text{NR}}\left(\psi_{l},\alpha_{l}\right) \triangleq -\mathbb{E}\left[\frac{\partial^{2} L(\mathbf{y})}{\partial \psi_{l} \partial \alpha_{l}}\right]$}
% Differentiating Eq. (\ref{eqn_fim_phi_1}) w.r.t $\alpha_{l}$, noting that $\frac{\partial \mathbf{m}^{H}}{\partial \alpha_{l}}=0$ and subsequently applying $-\mathbb{E}$ w.r.t $p\left(\mathbf{y}|(.),(.),(.)\right)$, we obtain
As in the computation of $\mathbf{J}_{\text{NR}}\left(\phi_{l},\alpha_{l}\right)$, it is straightforward to show
\begin{eqnarray}
\mathbf{J}_{\text{NR}}\left(\psi_{l},\alpha_{l}\right)&=&\frac{1}{\sigma_{v}^{2}} \left[\frac{\partial \mathbf{m}^{\dagger}}{\partial \psi_{l}}\frac{\partial \mathbf{m}}{\partial \alpha_{l}} + \frac{\partial \mathbf{m}^{\dagger}}{\partial \alpha_{l}}\frac{\partial \mathbf{m}}{\partial \psi_{l}}\right]\nonumber
\end{eqnarray}
Using Eq. (\ref{eqn_fim_phi_alpha_2}), Eq. (\ref{eqn_fim_phi_psi_2}) and Eq. (\ref{eqn_fim_phi_psi_3}) in the equation above, which upon further simplification using the property of $\text{tr}[.]$ and $\otimes$ given in Eq. (\ref{trace_op_property}) and Eq. (\ref{kron_prod_property}) can be simplified to Eq. (\ref{eqn_fim_psi_alpha_2})
\begin{eqnarray}
\mathbf{J}_{\text{NR}}\left(\psi_{l},\alpha_{l}\right)&=&\frac{2 \alpha_{l} \pi \sin(\psi_{l})}{\sigma_{v}^{2}} \text{Im}\left[\left(\mathbf{e}_{t}^{T}(\phi_l) \otimes \widetilde{\mathbf{e}}_{r}^{T}(\psi_l)\right)\mathbf{K} \left(\mathbf{e}_{t}^{*}(\phi_{l}) \otimes \mathbf{e}_{r}(\psi_{l})\right)\right]\label{eqn_fim_psi_alpha_1}\\
&=&\frac{2 \alpha_{l} \pi \sin(\psi_{l})}{\sigma_{v}^{2}} \text{Im}\left\{\text{tr}\left[\mathbf{K} \left(\mathbf{e}_{t}^{*}(\phi_{l}) \otimes \mathbf{e}_{r}(\psi_{l})\right) \left(\mathbf{e}_{t}^{T}(\phi_l) \otimes \widetilde{\mathbf{e}}_{r}^{T}(\psi_l)\right)\right]\right\}\nonumber\\
&=&\frac{2 \alpha_{l} \pi \sin(\psi_{l})}{\sigma_{v}^{2}} \text{Im}\left\{\text{tr}\left[\mathbf{K} \left(\mathbf{P}^{(4)} \otimes \mathbf{Q}^{(2)}\right)\right]\right\}\label{eqn_fim_psi_alpha_2}\\
\text{where}~\mathbf{P}^{(4)} &\triangleq& \mathbf{e}^{*}_{t}(\phi_{l})\mathbf{e}_{t}^{T}(\phi_{l})~,\label{p4_matrix_def}
\end{eqnarray}
$\mathbf{K} \triangleq \mathbf{A}^{\dagger} \mathbf{A}$ and $\mathbf{Q}^{(2)}$ is defined as in Eq. (\ref{q2_matrix_def}).

The general element in matrix $\mathbf{P}^{(4)}$ can be written as follows:
\begin{eqnarray}
\mathbf{P}^{(4)}(r,s)&=&\frac{1}{n_{t}}e^{j\pi(r-s)\cos(\phi_{l})}. \label{p4_general_element_def}
\end{eqnarray}
\subsection*{Calculation of: $\mathbf{J}_{\text{NR}}\left(\phi_{l},\alpha_{m}\right) \triangleq -\mathbb{E}\left[\frac{\partial^{2} L(\mathbf{y})}{\partial \phi_{l} \partial \alpha_{m}}\right]~(l \neq m)$}
% Differentiating Eq. (\ref{eqn_fim_phi_1}) w.r.t $\alpha_{m}$, noting that $\frac{\partial \mathbf{m}^{H}}{\partial \alpha_{m}}=0$ and subsequently applying $-\mathbb{E}$ w.r.t $p\left(\mathbf{y}|(.),(.),(.)\right)$, we obtain
As before it is straightforward to show that
\begin{eqnarray}
\mathbf{J}_{\text{NR}}\left(\phi_{l},\alpha_{m}\right)&=&\frac{1}{\sigma_{v}^{2}} \left[\frac{\partial \mathbf{m}^{\dagger}}{\partial \phi_{l}}\frac{\partial \mathbf{m}}{\partial \alpha_{m}} + \frac{\partial \mathbf{m}^{\dagger}}{\partial \alpha_{m}}\frac{\partial \mathbf{m}}{\partial \phi_{l}}\right]\label{eqn_fim_phi_alpha_diff_0}\\
&=&\frac{- 2 \pi \alpha_{l} \sin(\phi_{l})}{\sigma_{v}^{2}} \text{Im}\left[\left(\widetilde{\mathbf{e}}_{t}^{\dagger}(\phi_l) \otimes \mathbf{e}_{r}^{\dagger}(\psi_l)\right)\mathbf{K} \left(\mathbf{e}_{t}^{*}(\phi_{m}) \otimes \mathbf{e}_{r}(\psi_{m})\right)\right]\label{eqn_fim_phi_alpha_diff_1}\\
&=&\frac{- 2 \pi \alpha_{l} \sin(\phi_{l})}{\sigma_{v}^{2}} \text{Im}\left\{\text{tr}\left[\mathbf{K} \left(\mathbf{e}_{t}^{*}(\phi_{m}) \otimes \mathbf{e}_{r}(\psi_{m})\right)\left(\widetilde{\mathbf{e}}_{t}^{\dagger}(\phi_l) \otimes \mathbf{e}_{r}^{\dagger}(\psi_l)\right)\right]\right\}\label{eqn_fim_phi_alpha_diff_2}\\
&=&\frac{- 2 \pi \alpha_{l} \sin(\phi_{l})}{\sigma_{v}^{2}} \text{Im}\left\{\text{tr}\left[\mathbf{K} \left(\mathbf{P}^{(5)} \otimes \mathbf{Q}^{(3)}\right)\right]\right\}\label{eqn_fim_phi_alpha_diff_3}\\
\text{where}~ \mathbf{P}^{(5)} &\triangleq& \mathbf{e}^{*}_{t}(\phi_{m}) \sin(\phi_{l})\widetilde{\mathbf{e}}_{t}^{\dagger}(\phi_{l}),\label{p5_matrix_def}\\
\mathbf{Q}^{(3)} &\triangleq& \mathbf{e}_{r}(\psi_{m})\mathbf{e}_{r}^{\dagger}(\psi_{l}),\label{q3_matrix_def}
\end{eqnarray}
and $\mathbf{K} \triangleq \mathbf{A}^{\dagger}\mathbf{A}$. The Eq. (\ref{eqn_fim_phi_alpha_diff_1}) follows from Eq. (\ref{eqn_fim_phi_alpha_diff_0}) by using Eq. (\ref{eqn_fim_phi_4}), Eq. (\ref{eqn_fim_phi_5}) and Eq. (\ref{eqn_fim_phi_alpha_2}). The Eq. (\ref{eqn_fim_phi_alpha_diff_2}) and Eq. (\ref{eqn_fim_phi_alpha_diff_3}) follow from Eq. Eq. (\ref{eqn_fim_phi_alpha_diff_1}) by using the properties of $\text{tr}[.]$ and $\otimes$ operators respectively given in Eq. (\ref{trace_op_property}) and Eq. (\ref{kron_prod_property}) respectively.

The general element in matrix $\mathbf{P}^{(5)}$ and $\mathbf{Q}^{(3)}$ can be written as follows:
\begin{eqnarray}
\mathbf{P}^{(5)}(r,s)&=&\frac{(s-1)}{n_{t}}e^{j\pi(r-1)\cos(\phi_{m})} e^{-j\pi(s-1)\cos(\phi_{l})}\sin(\phi_{l})\label{p5_general_element_def}\\
\mathbf{Q}^{(3)}(r,s)&=&\frac{1}{n_{r}} e^{-j\pi(r-1)\cos(\psi_{m})} e^{j\pi(s-1)\cos(\psi_{l})}\label{q3_general_element_def}
\end{eqnarray}
\subsection*{Calculation of: $\mathbf{J}_{\text{NR}}\left(\psi_{l},\alpha_{m}\right) \triangleq -\mathbb{E}\left[\frac{\partial^{2} L(\mathbf{y})}{\partial \psi_{l} \partial \alpha_{m}}\right]~(l \neq m)$}
It is straightforward to show that
\begin{eqnarray}
\mathbf{J}_{\text{NR}}\left(\psi_{l},\alpha_{m}\right)&=&\frac{1}{\sigma_{v}^{2}} \left[\frac{\partial \mathbf{m}^{\dagger}}{\partial \psi_{l}}\frac{\partial \mathbf{m}}{\partial \alpha_{m}} + \frac{\partial \mathbf{m}^{\dagger}}{\partial \alpha_{m}}\frac{\partial \mathbf{m}}{\partial \psi_{l}}\right]\label{eqn_fim_psi_alpha_diff_0}
\end{eqnarray}

Using Eq. (\ref{eqn_fim_phi_alpha_2}), Eq. (\ref{eqn_fim_phi_psi_2}) and Eq. (\ref{eqn_fim_phi_psi_3}) in Eq. (\ref{eqn_fim_psi_alpha_diff_0}), and noting that $\mathbf{K} \triangleq \mathbf{A}^{\dagger}\mathbf{A}$, we obtain
\begin{eqnarray}
\mathbf{J}_{\text{NR}}\left(\psi_{l},\alpha_{m}\right) &=&\frac{2 \pi \alpha_{l} \sin(\psi_{l})}{\sigma_{v}^{2}} \text{Im}\left[ \left(\mathbf{e}_{t}^{T}(\phi_l) \otimes \widetilde{\mathbf{e}}_{r}^{T}(\psi_l)\right) \mathbf{K} \left(\mathbf{e}_{t}^{*}(\phi_{m}) \otimes \mathbf{e}_{r}(\psi_{m})\right)\right].\label{eqn_fim_psi_alpha_diff_1}
\end{eqnarray}

Using the properties of $\text{tr}[.]$ and $\otimes$ operators respectively given in Eq. (\ref{trace_op_property}) and Eq. (\ref{kron_prod_property}) respectively Eq. (\ref{eqn_fim_psi_alpha_diff_1}) can be simplified as follows:
\begin{eqnarray}
\mathbf{J}_{\text{NR}}\left(\psi_{l},\alpha_{m}\right) &=&\frac{2 \pi \alpha_{l} \sin(\psi_{l})}{\sigma_{v}^{2}} \text{Im}\left\{\text{tr}\left[\mathbf{K} \left(\mathbf{e}_{t}^{*}(\phi_{m}) \otimes \mathbf{e}_{r}(\psi_{m})\right) \left(\mathbf{e}_{t}^{T}(\phi_l) \otimes \widetilde{\mathbf{e}}_{r}^{T}(\psi_l)\right)\right]\right\}\nonumber\\
&=&\frac{2 \pi \alpha_{l}}{\sigma_{v}^{2}}\text{Im}\left\{\text{tr}\left[\mathbf{K} \left(\mathbf{P}^{(6)} \otimes \mathbf{Q}^{(4)}\right)\right]\right\},\label{eqn_fim_psi_alpha_diff_2}\\
\text{where}~\mathbf{P}^{(6)} &\triangleq& \mathbf{e}^{*}_{t}(\phi_{m}) \mathbf{e}_{t}^{T}(\phi_{l})=\left(\mathbf{e}_{t}(\phi_{m}) \mathbf{e}_{t}^{\dagger}(\phi_{l})\right)^{*}\label{p6_matrix_def}\\
\text{and}~\mathbf{Q}^{(4)} &\triangleq& \mathbf{e}_{r}(\psi_{m})\widetilde{\mathbf{e}}_{r}^{T}(\psi_{l}) \sin(\psi_{l}).\label{q4_matrix_def}
\end{eqnarray}
The form of $\mathbf{P}^{(6)}$ above is similar to the form of $\mathbf{Q}^{(3)}$, defined in the calculation of $\mathbf{J}_{\text{NR}}(\phi_{l},\alpha_{m})$. Inferring from the generic element of $\mathbf{Q}^{(3)}$, we have generic element of the matrix $(\mathbf{e}_{t}(\phi_{m}) \mathbf{e}_{t}^{\dagger}(\phi_{l}))$ to be the following: $\frac{1}{n_{t}}e^{-j\pi(r-1)\cos(\phi_{m})} e^{j\pi(s-1)\cos(\phi_{l})}$. Hence, we have the generic element of $\mathbf{P}^{(6)}$ to be the following:
\begin{eqnarray}
\mathbf{P}^{(6)}(r,s)&=&\frac{1}{n_{t}}\left(e^{-j\pi(r-1)\cos(\phi_{m})} \right)^{*}\left(e^{j\pi(s-1)\cos(\phi_{l})} \right)^{*}~,\nonumber\\
&=&\frac{1}{n_{t}}e^{j\pi(r-1)\cos(\phi_{m})} e^{-j\pi(s-1)\cos(\phi_{l})}~.\label{p6_general_element_def}
\end{eqnarray}

The generic element of $\mathbf{Q}^{(4)}$ can be written as follows:
\begin{equation}
\mathbf{Q}^{(4)}(r,s)=\frac{(s-1)}{n_{r}} e^{-j\pi(r-1)\cos(\psi_{m})} e^{j\pi(s-1)\cos(\psi_{l})} \sin(\psi_{l})~.\label{q4_general_element_def} 
\end{equation}
\subsection*{Calculation of: $\mathbf{J}_{\text{NR}}\left(\phi_{l},\phi_{m}\right) \triangleq -\mathbb{E}\left[\frac{\partial^{2} L(\mathbf{y})}{\partial \phi_{l} \partial \phi_{m}}\right]~(l \neq m)$}
Differentiating Eq. (\ref{eqn_fim_phi_2}) w.r.t $\phi_m$ and taking $-\mathbb{E}$ of the resulting we obtain
\begin{equation}
\mathbf{J}_{\text{NR}}\left(\phi_{l},\phi_{m}\right)=\frac{1}{\sigma_{v}^{2}} \left[\frac{\partial \mathbf{m}^{\dagger}}{\partial \phi_{l}}\frac{\partial \mathbf{m}}{\partial \phi_{m}} + \frac{\partial \mathbf{m}^{\dagger}}{\partial \phi_{m}}\frac{\partial \mathbf{m}}{\partial \phi_{l}}\right]\label{eqn_fim_phi_phi_diff_1}
\end{equation}
The expressions for all the four terms within $[.]$ in the equation above can be gleaned from Eq. (\ref{eqn_fim_phi_4}) and Eq. (\ref{eqn_fim_phi_5}). Further using $\mathbf{K} \triangleq \mathbf{A}^{\dagger}\mathbf{A}$, we obtain
\begin{eqnarray}
\mathbf{J}_{\text{NR}}\left(\phi_{l},\phi_{m}\right)&=&\frac{\pi^{2}\sin(\phi_{l})\sin(\phi_{m})}{\sigma_{v}^{2}}\left[\text{term 1}~ + \text{term 2}\right]\nonumber\\
\text{where term 1}&=&\alpha_{l}\alpha_{m}\left(\widetilde{\mathbf{e}}_{t}^{\dagger}(\phi_{l}) \otimes \mathbf{e}^{\dagger}_{r}(\psi_l)\right)\mathbf{K}\left(\widetilde{\mathbf{e}}_{t}(\phi_{m}) \otimes \mathbf{e}_{r}(\psi_m)\right),\nonumber\\
\text{and term 2}&=&\alpha_{l}\alpha_{m}\left(\widetilde{\mathbf{e}}_{t}^{\dagger}(\phi_{m}) \otimes \mathbf{e}^{\dagger}_{r}(\psi_m)\right)\mathbf{K}\left(\widetilde{\mathbf{e}}_{t}(\phi_{l}) \otimes \mathbf{e}_{r}(\psi_l)\right)\nonumber\\
&=&\frac{2 \pi^{2}\sin(\phi_{l})\sin(\phi_{m}) \alpha_{l}\alpha_{m}}{\sigma_{v}^{2}}~\text{Re}\left\{\left(\widetilde{\mathbf{e}}_{t}^{\dagger}(\phi_{l}) \otimes \mathbf{e}^{\dagger}_{r}(\psi_l)\right)\mathbf{K}\left(\tilde{\mathbf{e}}_{t}(\phi_{m}) \otimes \mathbf{e}_{r}(\psi_m)\right)\right\}.\label{eqn_fim_phi_phi_diff_2}
\end{eqnarray}

Using the properties of $\text{tr}[.]$ and $\otimes$ operators respectively given in Eq. (\ref{trace_op_property}) and Eq. (\ref{kron_prod_property}) respectively, we can simplify Eq. (\ref{eqn_fim_phi_phi_diff_2}) as follows:
\begin{eqnarray}
\mathbf{J}_{\text{NR}}\left(\phi_{l},\phi_{m}\right)&=&\frac{2 \pi^{2}\sin(\phi_{l})\sin(\phi_{m}) \alpha_{l}\alpha_{m}}{\sigma_{v}^{2}}~\text{Re}\left\{\text{tr}\left[\mathbf{K}\left(\widetilde{\mathbf{e}}_{t}(\phi_{m}) \otimes \mathbf{e}_{r}(\psi_m)\right)(\widetilde{\mathbf{e}}_{t}^{\dagger}(\phi_{l}) \otimes \mathbf{e}^{\dagger}_{r}(\psi_l))\right]\right\}\nonumber\\
&=&\frac{2 \pi^{2} \alpha_{l}\alpha_{m}}{\sigma_{v}^{2}}\text{Re}\left\{\text{tr}\left[\mathbf{K} \left(\mathbf{P}^{(7)} \otimes \mathbf{Q}^{(3)}\right)\right]\right\},\label{eqn_fim_phi_phi_diff_3}\\
\text{where}~\mathbf{P}^{(7)} &\triangleq& \widetilde{\mathbf{e}}_{t}(\phi_{m}) \widetilde{\mathbf{e}}_{t}^{\dagger}(\phi_{l}) \sin(\phi_{m}) \sin(\phi_{l})~,\label{p7_matrix_def}
\end{eqnarray}
and $\mathbf{Q}^{(3)}$ is defined as in Eq. (\ref{q3_matrix_def}).

The generic element of matrix $\mathbf{P}^{(7)}$ can be expressed and simplified further to the form given below:
\begin{eqnarray}
\mathbf{P}^{(7)}(r,s) &=& \frac{(r-1)(s-1)}{n_{t}} e^{j \pi (r-1) \cos(\phi_{m})} \sin(\phi_{m})e^{-j \pi (s-1) \cos(\phi_{l})} \sin(\phi_{l})\label{p7_general_element_def}
\end{eqnarray}
\subsection*{Calculation of: $\mathbf{J}_{\text{NR}}\left(\psi_{l},\psi_{m}\right) \triangleq -\mathbb{E}\left[\frac{\partial^{2} L(\mathbf{y})}{\partial \psi_{l} \partial \psi_{m}}\right]~(l \neq m)$}
As in the calculation of $\mathbf{J}_{\text{NR}}\left(\phi_{l},\phi_{m}\right)$ immediately above, one can show 
\begin{equation}
\mathbf{J}_{\text{NR}}\left(\psi_{l},\psi_{m}\right)=\frac{1}{\sigma_{v}^{2}} \left[\frac{\partial \mathbf{m}^{\dagger}}{\partial \psi_{l}}\frac{\partial \mathbf{m}}{\partial \psi_{m}} + \frac{\partial \mathbf{m}^{\dagger}}{\partial \psi_{m}}\frac{\partial \mathbf{m}}{\partial \psi_{l}}\right]\label{eqn_fim_psi_psi_diff_1}
\end{equation}
The expressions for all the four terms within $[.]$ in the equation above can be gleaned from Eq. (\ref{eqn_fim_phi_psi_2}) and Eq. (\ref{eqn_fim_phi_psi_3}). Further using $\mathbf{K} \triangleq \mathbf{A}^{\dagger}\mathbf{A}$, we obtain
\begin{eqnarray}
\mathbf{J}_{\text{NR}}\left(\psi_{l},\psi_{m}\right)&=&\frac{\pi^{2}\sin(\psi_{l})\sin(\psi_{m})}{\sigma_{v}^{2}}\left[\text{term 1}~ + \text{term 2}\right]\nonumber\\
\text{where term 1}&=&\alpha_{l}\alpha_{m}\left(\mathbf{e}_{t}^{T}(\phi_{m}) \otimes \widetilde{\mathbf{e}}^{T}_{r}(\psi_m)\right)\mathbf{K}\left(\mathbf{e}_{t}^{*}(\phi_{l}) \otimes \widetilde{\mathbf{e}}^{*}_{r}(\psi_l)\right),\nonumber\\
\text{and term 2}&=&\alpha_{l}\alpha_{m}\left(\mathbf{e}_{t}^{T}(\phi_{l}) \otimes \widetilde{\mathbf{e}}^{T}_{r}(\psi_l)\right)\mathbf{K}\left(\mathbf{e}_{t}^{*}(\phi_{m}) \otimes \tilde{\mathbf{e}}^{*}_{r}(\psi_m)\right)\nonumber\\
&=&\frac{2 \pi^{2}\sin(\psi_{l})\sin(\psi_{m}) \alpha_{l}\alpha_{m}}{\sigma_{v}^{2}}~\text{Re}\left\{\left(\mathbf{e}_{t}^{T}(\phi_{m}) \otimes \widetilde{\mathbf{e}}^{T}_{r}(\psi_m)\right)\mathbf{K}\left(\mathbf{e}_{t}^{*}(\phi_{l}) \otimes \widetilde{\mathbf{e}}^{*}_{r}(\psi_l)\right)\right\}.\label{eqn_fim_psi_psi_diff_2}
\end{eqnarray}

The Eq. (\ref{eqn_fim_psi_psi_diff_2}) can further be simplified using the properties of $\text{tr}[.]$ and $\otimes$ operators respectively given in Eq. (\ref{trace_op_property}) and Eq. (\ref{kron_prod_property}) respectively as follows:
\begin{eqnarray}
\mathbf{J}_{\text{NR}}\left(\psi_{l},\psi_{m}\right)&=&\frac{2 \pi^{2}\sin(\psi_{l})\sin(\psi_{m}) \alpha_{l}\alpha_{m}}{\sigma_{v}^{2}}~\text{Re}\left\{\text{tr}\left[\mathbf{K}(\mathbf{e}_{t}^{*}(\phi_{l}) \otimes \widetilde{\mathbf{e}}^{*}_{r}(\psi_l))(\mathbf{e}_{t}^{T}(\phi_{m}) \otimes \widetilde{\mathbf{e}}^{T}_{r}(\psi_m))\right]\right\}~,\nonumber\\
&=&\frac{2 \pi^{2} \alpha_{l} \alpha_{m}}{\sigma_{v}^{2}}\text{Re}\left\{\text{tr}\left[\mathbf{K} \left(\mathbf{P}^{(8)} \otimes \mathbf{Q}^{(5)}\right)\right]\right\},\label{eqn_fim_psi_psi_diff_3}\\
\text{where}~\mathbf{P}^{(8)} &\triangleq& \mathbf{e}^{*}_{t}(\phi_{l}) \mathbf{e}_{t}^{T}(\phi_{m})\label{p8_matrix_def}\\
\text{and}~ \mathbf{Q}^{(5)}&=&\widetilde{\mathbf{e}}^{*}_{r}(\psi_{l}) \widetilde{\mathbf{e}}_{r}^{T}(\psi_{m}) \sin(\psi_{l}) \sin(\psi_{m}) \label{q5_matrix_def}
\end{eqnarray}
The generic element of $\mathbf{P}^{(8)}$ and $\mathbf{Q}^{(5)}$ can be expressed as follows:
\begin{eqnarray}
\mathbf{P}^{(8)}(r,s)&=&\frac{1}{n_{t}} e^{j \pi (r-1) \cos(\phi_{l})} e^{-j \pi (s-1) \cos(\phi_{m})}~, \label{p8_general_element_def}\\
\mathbf{Q}^{(5)}(r,s)&=&\frac{(r-1)(s-1)}{n_{r}} e^{-j \pi (r-1) \cos(\psi_{l})} \sin(\psi_{l})e^{j \pi (s-1) \cos(\psi_{m})} \sin(\psi_{m})~.\label{q5_general_element_def}
\end{eqnarray}
\subsection*{Calculation of: $\mathbf{J}_{\text{NR}}\left(\psi_{l},\psi_{l}\right) \triangleq -\mathbb{E}\left[\frac{\partial^{2} L(\mathbf{y})}{\partial \psi_{l}^{2}}\right]$}
Setting $l=m$ in Eq. (\ref{eqn_fim_psi_psi_diff_2}) we obtain 
\begin{equation}
\mathbf{J}_{\text{NR}}\left(\psi_{l},\psi_{l}\right)= \frac{2 \pi^{2}\sin^{2}(\psi_{l}) \alpha_{l}^{2}}{\sigma_{v}^{2}}\left(\mathbf{e}_{t}^{T}(\phi_{l}) \otimes \widetilde{\mathbf{e}}^{T}_{r}(\psi_l)\right)\mathbf{K}\left(\mathbf{e}_{t}^{*}(\phi_{l}) \otimes \widetilde{\mathbf{e}}^{*}_{r}(\psi_l)\right).\label{eqn_fim_psi_1}
\end{equation}

Using the properties of $\text{tr}[.]$ and $\otimes$ operators respectively given in Eq. (\ref{trace_op_property}) and Eq. (\ref{kron_prod_property}) respectively we can simplify Eq. (\ref{eqn_fim_psi_1}) as follows:
\begin{eqnarray}
\mathbf{J}_{\text{NR}}\left(\psi_{l},\psi_{l}\right)&=&\frac{2 \pi^{2}\sin^{2}(\psi_{l}) \alpha_{l}^{2}}{\sigma_{v}^{2}}\left[\mathbf{K}(\mathbf{e}_{t}^{*}(\phi_{l}) \otimes \widetilde{\mathbf{e}}^{*}_{r}(\psi_l))\mathbf{e}_{t}^{T}(\phi_{l}) \otimes \widetilde{\mathbf{e}}^{T}_{r}(\psi_l)\right]\nonumber\\
&=&\frac{2 \pi^2 \alpha_{l}^{2}}{\sigma_{v}^{2}}\text{tr}\left[\mathbf{K} \left(\mathbf{P}^{(4)} \otimes \mathbf{Q}^{(6)}\right) \right]~,\label{eqn_fim_psi_2}\\
\text{where}~\mathbf{Q}^{(6)} &\triangleq& \widetilde{\mathbf{e}}^{*}_{r}(\psi_{l}) \widetilde{\mathbf{e}}^{T}_{r}(\psi_{l}) \sin^{2}(\psi_{l})=\left(\widetilde{\mathbf{e}}_{r}(\psi_{l}) \widetilde{\mathbf{e}}^{\dagger}_{r}(\psi_{l}) \sin^{2}(\psi_{l})\right)^{*}~,\label{q6_matrix_def}
\end{eqnarray}
and $\mathbf{P}^{(4)}$ is defined as in Eq. (\ref{p4_matrix_def}).

The form of $\mathbf{Q}^{(6)}$ above is similar to the form of $\mathbf{P}$, defined in the calculation of $\mathbf{J}_{\text{NR}}\left(\phi_{l},\phi_{l}\right)$. Inferring from the generic element of $\mathbf{P}$, we have generic element of the matrix $\left(\widetilde{\mathbf{e}}_{r}(\psi_{l}) \widetilde{\mathbf{e}}_{r}^{\dagger}(\psi_{l}) \sin^{2}(\psi_{l})\right)$ to be the following: $\frac{(r-1)(s-1)}{n_{r}}e^{-j\pi(s-r)\cos(\psi_{l})} \sin^{2}(\psi_{l})$. Hence, we have the generic element of $\mathbf{Q}^{(6)}$ to be the following:
\begin{eqnarray}
\mathbf{Q}^{(6)}(r,s)&=&\frac{(r-1)(s-1)}{n_{r}}\left(e^{-j\pi(s-r)\cos(\psi_{l})} \sin^{2}(\psi_{l})\right)^{*}\nonumber\\
&=&\frac{(r-1)(s-1)}{n_{r}}e^{j\pi(s-r)\cos(\psi_{l})} \sin^{2}(\psi_{l})\label{q6_general_element_def}
\end{eqnarray}
\subsection*{Calculation of: $\mathbf{J}_{\text{NR}}\left(\phi_{l},\psi_{m}\right) \triangleq -\mathbb{E}\left[\frac{\partial^{2} L(\mathbf{y})}{\partial \phi_{l}\psi_{m}}\right]$}
Differentiating Eq. (\ref{eqn_fim_phi_1}) w.r.t $\psi_m$ we obtain
\begin{equation}
\mathbf{J}_{\text{NR}}\left(\phi_{l},\psi_{m}\right)=\frac{1}{\sigma_{v}^{2}} \left[\frac{\partial \mathbf{m}^{\dagger}}{\partial \psi_{m}}\frac{\partial \mathbf{m}}{\partial \phi_{l}} + \frac{\partial \mathbf{m}^{\dagger}}{\partial \phi_{l}}\frac{\partial \mathbf{m}}{\partial \psi_{m}}\right]\label{eqn_fim_phi_psi_diff_1}
\end{equation}
The expressions for all the four terms within $[.]$ in the equation above can be gleaned from Eq. (\ref{eqn_fim_phi_4}), Eq. (\ref{eqn_fim_phi_5}), Eq. (\ref{eqn_fim_phi_psi_2}) and Eq. (\ref{eqn_fim_phi_psi_3}) respectively. Further noting that $\mathbf{K} \triangleq \mathbf{A}^{\dagger}\mathbf{A}$, we obtain
\begin{eqnarray}
\mathbf{J}_{\text{NR}}\left(\phi_{l},\psi_{m}\right)&=&\frac{-\pi^{2}\sin(\phi_{l})\sin(\psi_{m})}{\sigma_{v}^{2}}\left[\text{term 1} + \text{term 2}\right]\nonumber\\
&&\text{term 1}=\alpha_{l}\alpha_{m}\left(\mathbf{e}_{t}^{T}(\phi_{m}) \otimes \widetilde{\mathbf{e}}^{T}_{r}(\psi_m)\right)\mathbf{K}\left(\widetilde{\mathbf{e}}_{t}(\phi_{l}) \otimes \mathbf{e}_{r}(\psi_l)\right)\nonumber\\
&&\text{term 2}=\alpha_{l}\alpha_{m}\left(\tilde{\mathbf{e}}_{t}^{\dagger}(\phi_{l}) \otimes \mathbf{e}^{\dagger}_{r}(\psi_l)\right)\mathbf{K}\left(\mathbf{e}^{*}_{t}(\phi_{m}) \otimes \tilde{\mathbf{e}}^{*}_{r}(\psi_m)\right)\nonumber\\
&&=\frac{-2 \pi^{2}\sin(\phi_{l})\sin(\psi_{m})}{\sigma_{v}^{2}}~\text{Re}\left\{\alpha_{l}\alpha_{m}\left(\mathbf{e}_{t}^{T}(\phi_{m}) \otimes \widetilde{\mathbf{e}}^{T}_{r}(\psi_m)\right)\mathbf{K}\left(\widetilde{\mathbf{e}}_{t}(\phi_{l}) \otimes \mathbf{e}_{r}(\psi_l)\right)\right\}.\label{eqn_fim_phi_psi_diff_2}
\end{eqnarray}
Using the properties of $\text{tr}[.]$ and $\otimes$ operators respectively given in Eq. (\ref{trace_op_property}) and Eq. (\ref{kron_prod_property}) respectively, we can simplify Eq. (\ref{eqn_fim_phi_psi_diff_2}) as follows
\begin{eqnarray}
\mathbf{J}_{\text{NR}}\left(\phi_{l},\psi_{m}\right)&=&\frac{-2 \pi^{2}\sin(\phi_{l})\sin(\psi_{m})\alpha_{l}\alpha_{m}}{\sigma_{v}^{2}}~\text{Re}\left\{\text{tr}\left[(\mathbf{K}(\widetilde{\mathbf{e}}_{t}(\phi_{l}) \otimes \mathbf{e}_{r}(\psi_l)) (\mathbf{e}_{t}^{T}(\phi_{m}) \otimes \widetilde{\mathbf{e}}^{T}_{r}(\psi_m))\right]\right\}\nonumber\\
&=&\frac{-2 \pi^{2}\sin(\phi_{l})\sin(\psi_{m})\alpha_{l}\alpha_{m}}{\sigma_{v}^{2}}\text{Re}\left\{\text{tr}\left[\mathbf{K} \left(\mathbf{P}^{(9)} \otimes \mathbf{Q}^{(7)}\right)\right]\right\},\label{eqn_fim_phi_psi_diff_3}\\
\text{where}~\mathbf{P}^{(9)} &\triangleq& \tilde{\mathbf{e}}_{t}(\phi_{l}) \mathbf{e}_{t}^{T}(\phi_{m}) \sin(\phi_{l})~,\label{p9_matrix_def}\\
\text{and}~ \mathbf{Q}^{(7)} &\triangleq& \mathbf{e}_{r}(\psi_{l}) \tilde{\mathbf{e}}_{r}^{T}(\psi_{m}) \sin(\psi_{m})~. \label{q7_matrix_def}
\end{eqnarray}
The generic element of $\mathbf{P}^{(9)}$ and $\mathbf{Q}^{(7)}$ can be expressed as follows:
\begin{eqnarray}
\mathbf{P}^{(9)}(r,s)&=&\frac{(r-1)}{n_{t}} e^{j \pi (r-1) \cos(\phi_{l})} \sin(\phi_{l})e^{-j \pi (s-1) \cos(\phi_{m})}~, \label{p9_general_element_def}\\
\mathbf{Q}^{(7)}(r,s)&=&\frac{(s-1)}{n_{r}} e^{-j \pi (r-1) \cos(\psi_{l})}e^{j \pi (s-1) \cos(\psi_{m})} \sin(\psi_{m})~.\label{q7_general_element_def}
\end{eqnarray}
\subsection*{Calculation of: $\mathbf{J}_{\text{NR}}\left(\alpha_{l},\alpha_{l}\right) \triangleq -\mathbb{E}\left[\frac{\partial^{2} L(\mathbf{y})}{\partial \alpha_{l}^{2}}\right]$}
Differentiating Eq. (\ref{llr_eqn}) w.r.t $\alpha_l$ and applying $\mathbb{E}$ operator w.r.t the PDF $p\left(\mathbf{y}|\boldsymbol{\theta}\right)$ we obtain
\begin{equation}
\mathbf{J}_{\text{NR}}\left(\alpha_{l},\alpha_{l}\right)=\frac{2}{\sigma_{v}^{2}} \frac{\partial \mathbf{m}^{\dagger}}{\partial \alpha_{l}}\frac{\partial \mathbf{m}}{\partial \alpha_{l}}\label{eqn_fim_alpha_1}
\end{equation}
Using Eq. (\ref{eqn_fim_phi_alpha_2}) in Eq. (\ref{eqn_fim_alpha_1}) and noting that $\mathbf{K} \triangleq \mathbf{A}^{\dagger}\mathbf{A}$, we obtain
\begin{equation}
\mathbf{J}_{\text{NR}}\left(\alpha_{l},\alpha_{l}\right)=\frac{2}{\sigma_{v}^{2}}\left(\mathbf{e}_{t}^{T}(\phi_{l}) \otimes \mathbf{e}_{r}^{\dagger}(\psi_{l})\right) \mathbf{K} \left(\mathbf{e}_{t}^{*}(\phi_{l}) \otimes \mathbf{e}_{r}(\psi_{l})\right)\label{eqn_fim_alpha_2}
\end{equation}
Using the properties of $\text{tr}[.]$ and $\otimes$ operators respectively given in Eq. (\ref{trace_op_property}) and Eq. (\ref{kron_prod_property}) respectively, we can simplify Eq. (\ref{eqn_fim_alpha_2}) as follows
\begin{eqnarray}
\mathbf{J}_{\text{NR}}\left(\alpha_{l},\alpha_{l}\right)&=&\frac{2}{\sigma_{v}^{2}}\text{tr}\left[\mathbf{K}(\mathbf{e}_{t}^{*}(\phi_{l}) \otimes \mathbf{e}_{r}(\psi_{l}))(\mathbf{e}_{t}^{T}(\phi_{l}) \otimes \mathbf{e}_{r}^{\dagger}(\psi_{l}))\right]\nonumber\\
&=&\frac{2}{\sigma_{v}^{2}}\text{tr}\left[\mathbf{K} \left(\mathbf{P}^{(4)} \otimes \mathbf{Q}\right)\right]~,\label{eqn_fim_alpha_3}
\end{eqnarray}
where $\mathbf{Q}$ and $\mathbf{P}^{(4)}$ are defined as in Eq. (\ref{pq_matrix_def}) and Eq. (\ref{p4_matrix_def}).
\subsection*{Calculation of: $\mathbf{J}_{\text{NR}}\left(\alpha_{l},\alpha_{m}\right) \triangleq -\mathbb{E}\left[\frac{\partial^{2} L(\mathbf{y})}{\partial \alpha_{l} \partial \alpha_{m}}\right]$}
Differentiating Eq. (\ref{llr_eqn}) w.r.t $\alpha_l$ and $\alpha_{m}$ successively and applying $\mathbb{E}$ operator w.r.t the PDF $p\left(\mathbf{y}|\mathbf{\theta}\right)$ we obtain
\begin{equation}
\mathbf{J}_{\text{NR}}\left(\alpha_{l},\alpha_{m}\right)=\frac{1}{\sigma_{v}^{2}} \left[\frac{\partial \mathbf{m}^{\dagger}}{\partial \alpha_{l}}\frac{\partial \mathbf{m}}{\partial \alpha_{m}} + \frac{\partial \mathbf{m}^{\dagger}}{\partial \alpha_{m}}\frac{\partial \mathbf{m}}{\partial \alpha_{l}}\right]\label{eqn_fim_alpha_alpha_diff_1}
\end{equation}
Using Eq. (\ref{eqn_fim_phi_alpha_2}) in Eq. (\ref{eqn_fim_alpha_alpha_diff_1}) and noting that $\mathbf{K} \triangleq \mathbf{A}^{\dagger}\mathbf{A}$, we obtain
\begin{equation}
\mathbf{J}_{\text{NR}}\left(\alpha_{l},\alpha_{m}\right)=\frac{2}{\sigma_{v}^{2}}\text{Re}\left\{(\mathbf{e}_{t}^{T}(\phi_{l}) \otimes \mathbf{e}_{r}^{\dagger}(\psi_{l})) \mathbf{K} (\mathbf{e}_{t}^{*}(\phi_{m}) \otimes \mathbf{e}_{r}(\psi_{m}))\right\}~.\label{eqn_fim_alpha_alpha_diff_2}
\end{equation}
Using the properties of $\text{tr}[.]$ and $\otimes$ operators respectively given in Eq. (\ref{trace_op_property}) and Eq. (\ref{kron_prod_property}) respectively, we can simplify Eq. (\ref{eqn_fim_alpha_alpha_diff_2}) as follows
\begin{eqnarray}
\mathbf{J}_{\text{NR}}\left(\alpha_{l},\alpha_{m}\right)&=&\frac{2}{\sigma_{v}^{2}}\text{Re}\left\{\text{tr}\left[\mathbf{K} (\mathbf{e}_{t}^{*}(\phi_{m}) \otimes \mathbf{e}_{r}(\psi_{m}))(\mathbf{e}_{t}^{T}(\phi_{l}) \otimes \mathbf{e}_{r}^{\dagger}(\psi_{l}))\right]\right\}~,\nonumber\\
&=&\frac{2}{\sigma_{v}^{2}}\text{Re}\left\{\text{tr}\left[\mathbf{K} \left(\mathbf{P}^{(6)} \otimes \mathbf{Q}^{(3)}\right)\right]\right\},\label{eqn_fim_alpha_alpha_diff_3}
\end{eqnarray}
where $\mathbf{Q}^{(3)}$ and $\mathbf{P}^{(6)}$ are defined as in Eq. (\ref{q3_matrix_def}) and Eq. (\ref{p6_matrix_def}).

\section*{Appendix B: Calculation of various entries of $\widetilde{\mathbf{J}}_{\text{P}}$}
As noted in Section \ref{subsect:bayesian_crlb} the Bayesian FIM $\widetilde{\mathbf{J}}_{\text{B}}$ contains two parts namely: $\widetilde{\mathbf{J}}_{\text{P}}$ and $\widetilde{\mathbf{J}}_{\text{D}}$. In this section we shall compute the entries of $\widetilde{\mathbf{J}}_{\text{P}}$.

We assumed $\{\phi_{l}\}~\text{and}~\{\psi_{l}\}~(l=1, \cdots, L)$ are uniformly distributed over $[0,\pi]$ and $\alpha_{l}~(l=1,\cdots,L)$ to be of Rician distribution. Further all the parameters are assumed to be independent of each other, therefore we have 
\begin{equation}
p(\boldsymbol{\theta}) = \left(\frac{1}{\pi}\right)^{2L} \Pi_{l=1}^{L}p_{\alpha}(\alpha_{l})~,\label{prior_par_pdf}
\end{equation}
where $p_{\alpha}(\alpha)=\frac{2\left(K + 1 \right)}{\Omega}\alpha e^{-\frac{(K + 1)}{\Omega}\left(\alpha^{2} + \frac{\Omega K}{K + 1}\right)}I_{0}\left(2\alpha \sqrt{\frac{K(K + 1)}{\Omega}}\right)$ denotes a Rician distribution with parameters $K$ (Rice factor) and $\Omega$ (second moment) \cite{PS2008}.

In general the $(i,j)^{\text{th}}$ entry of the matrix $\widetilde{\mathbf{J}}_{\text{P}}$ is given by $-\mathbb{E}\left[\frac{\partial^{2}\ln(p(\boldsymbol{\theta}))}{\partial \theta_{i} \partial \theta_{j}}\right]$ ($\mathbb{E}$ is w.r.t the PDF of $\boldsymbol{\theta}$ i.e. $p(\boldsymbol{\theta})$). Since $p(\boldsymbol{\theta})$ is independent of $\phi_{l}$ and $\psi_{l}$, therefore the following terms for any $l$ and $m$ would be zero: $\widetilde{\mathbf{J}}_{\text{P}}(\phi_{l},\psi_{m}) \triangleq -\mathbb{E}\left[\frac{\partial^{2}\ln(p(\boldsymbol{\theta}))}{\partial \phi_{l} \partial \phi_{m}}\right]$, $\widetilde{\mathbf{J}}_{\text{P}}(\phi_{l},\alpha_{m}) \triangleq -\mathbb{E}\left[\frac{\partial^{2}\ln(p(\boldsymbol{\theta}))}{\partial \phi_{l} \partial \alpha_{m}}\right]$ and $\widetilde{\mathbf{J}}_{\text{P}}(\psi_{l},\alpha_{m}) \triangleq -\mathbb{E}\left[\frac{\partial^{2}\ln(p(\boldsymbol{\theta}))}{\partial \psi_{l} \partial \alpha_{m}}\right]$. Further for $l \neq m$, $\widetilde{\mathbf{J}}_{\text{P}}(\alpha_{l},\alpha_{m}) \triangleq -\mathbb{E}\left[\frac{\partial^{2}\ln(p(\boldsymbol{\theta}))}{\partial \alpha_{l} \partial \alpha_{m}}\right]$ will also be equal to $0$.

From the above one can conclude that the matrix $\widetilde{\mathbf{J}}_{\text{P}}$ would be all zero except for the following $L$ terms. These $L$ terms are $\widetilde{\mathbf{J}}_{\text{P}}(\alpha_{l},\alpha_{l}) \triangleq -\mathbb{E}\left[\frac{\partial^{2}\ln(p(\boldsymbol{\theta}))}{\partial \alpha_{l}^{2}}\right]~(l=1,\cdots, L)$. It is a straightforward exercise to show that
\begin{eqnarray}
\frac{\partial^{2}\ln(p(\boldsymbol{\theta}))}{\partial \alpha_{l}^{2}}&=&\frac{-1}{\alpha_{l}^{2}} - \frac{2\left(K_{l} + 1\right)}{\Omega_{l}} + \frac{2 K_{l}\left(K_{l} + 1\right)}{\Omega_{l}}\left[ 1 + \text{term 1} + \text{term 2}\right]\label{eqn_fim_prior_mtx_1}\\
\text{where}~\text{term 1}&=&\frac{-J_{2}\left(-2j\alpha_{l}\sqrt{\frac{K_{l}(K_{l} + 1)}{\Omega_{l}}}\right)}{I_{0}\left(2 \alpha_{l}\sqrt{\frac{K_{l}(K_{l} + 1)}{\Omega_{l}}}\right)}\nonumber\\
\text{and}~\text{term 2}&=&\frac{2 \left(J_{1}\left(-2j\alpha_{l}\sqrt{\frac{K_{l}(K_{l} + 1)}{\Omega_{l}}}\right)\right)^{2}}{\left(I_{0}\left(2 \alpha_{l}\sqrt{\frac{K_{l}(K_{l} + 1)}{\Omega_{l}}}\right)\right)^{2}}~.\nonumber
\end{eqnarray}
In the last two equations above $K_{l}$ denotes the Rice factor of path $l$ and $\Omega_{l}$ denotes the second moment of path $l$ which is defined as in Section \ref{subsect:numresults_description}.
For each $l=1,\cdots, L$, $\widetilde{\mathbf{J}}_{\text{P}}(\alpha_{l},\alpha_{l})$ can be computed by taking $\mathbb{E}$ of the term given in Eq. (\ref{eqn_fim_prior_mtx_1}) w.r.t the PDF of $\boldsymbol{\theta}$ i.e. $p(\boldsymbol{\theta})$ (since all the components of $\boldsymbol{\theta}$ are independent of each other this is equivalent to simply taking $\mathbb{E}$ w.r.t the PDF of $\alpha_{l}$ i.e. $p_{\alpha}(\alpha_{l})$).
\section*{Appendix C: Calculation of various entries of $\widetilde{\mathbf{J}}_{\text{D}}$}
In general the $(i,j)^{\text{th}}$ entry of the matrix $\widetilde{\mathbf{J}}_{\text{D}}$ is given by $-\mathbb{E}\left[\frac{\partial^{2}L(\mathbf{y})}{\partial \theta_{i} \partial \theta_{j}}\right]$ ($\mathbb{E}$ is w.r.t the joint PDF of $\mathbf{y}$ and $\boldsymbol{\theta}$ i.e. $p(\mathbf{y},\boldsymbol{\theta})$). Since the $(i,j)^{\text{th}}$ entry of $\mathbf{J}_{\text{NR}}$ is obtained by averaging $\frac{\partial^{2}L(\mathbf{y})}{\partial \theta_{i} \partial \theta_{j}}$ w.r.t the PDF of $(\mathbf{y}|\mathbf{\theta})$, therefore $(i,j)^{\text{th}}$ entry of $\widetilde{\mathbf{J}}_{\text{D}}$ can be obtained by averaging the corresponding entry of $\mathbf{J}_{\text{NR}}$ w.r.t the PDF of $\boldsymbol{\theta}$. Using this fact we shall compute the various entries of $\widetilde{\mathbf{J}}_{\text{D}}$ in the ensuing calculations.

Before we proceed to the actual calculations we shall state a few well known integral expressions and their closed forms which would be useful for all the calculations in this section.

Consider the following integral
\begin{eqnarray}
&&\frac{1}{\pi}\int_{0}^{\pi}\sin^{2}(\theta) e^{jz \cos(\theta)} d\theta\nonumber\\
&=&\frac{1}{\pi}\int_{0}^{\pi}\left(\frac{1 - \cos(2 \theta)}{2}\right) e^{jz \cos(\theta)} d\theta\nonumber\\
&=&\frac{1}{\pi}\int_{0}^{\pi}\frac{1}{2} e^{jz\cos(\theta)} d\theta - \frac{1}{\pi}\int_{0}^{\pi}\frac{1}{2}\cos(2\theta) e^{jz\cos(\theta)} d\theta \nonumber\\
&=&\frac{1}{2}J_{0}(z) - \frac{1}{2}\frac{J_{2}(z)}{j^{-2}}\nonumber\\
&=&\frac{1}{2}\left[J_{0}(z) + J_{2}(z)\right]~,\label{bessel_fn_integral_1}
\end{eqnarray}
where the second last equation follows from the following definition of an $n^{\text{th}}$ order Bessel function of first kind: $J_{n}(z) \triangleq \frac{1}{\pi}\int_{0}^{\pi} e^{jz \cos(\theta)} \cos(n \theta) d\theta$.

Next consider the following integral
\begin{eqnarray}
&&\int_{0}^{\pi} e^{j \pi z \cos(\theta)} \sin(\theta) d\theta = \left(\frac{2}{z}\right)^{1/2}J_{1/2}(z)~,\label{bessel_fn_integral_2}
\end{eqnarray}
where the above equation follows from the the following result from Gradshteyn and Ryhik's book on integral tables (pg. 486 Sect. 3.915, Eq. (5)): $\int_{0}^{\pi} e^{j \beta \cos(x)} \sin^{2 \nu} (x) dx = \sqrt{\pi}\left(\frac{2}{\beta}\right)^{\nu} \Gamma\left(\nu + \frac{1}{2}\right) J_{\nu}(\beta)$ $\left(\text{Re}(\nu) > -\frac{1}{2}\right)$.
% The Bayesian FIM is obtained by averaging the various terms in FIM w.r.t the distribution of $\phi_{l}$, $\psi_{l}$ and $\alpha_{l}$. We assume the fading process across various paths to be according to the Ricean model. Therefore, accordingly the first and second moments of the gain $\alpha_l$ on the $l^{\text{th}}$ path is given by:
% \begin{eqnarray}
% \mathbb{E}\left[\alpha_{l}\right]&=&\sigma_{l}\sqrt{\frac{\pi}{2}} e^{-\frac{\kappa_{l}}{2}}\left[\left(1 + \kappa_{l}\right)I_{0}\left(\frac{\kappa_{l}}{2}\right) + \kappa_{l} I_{1}\left(\frac{\kappa_{l}}{2}\right)\right]~,\label{mean_alphal_def}\\
% \mathbb{E}\left[\alpha_{l}^{2}\right]&=& 2 \sigma_{l}^{2} + s_{l}^{2}~,\label{second_moment_alphal_def}
% \end{eqnarray}
% where $\kappa_{l}=\frac{s_{l}^{2}}{2 \sigma_{l}^{2}}$.
\subsection*{Calculation of $\widetilde{\mathbf{J}}_{\text{D}}(\phi_{l},\phi_{l}) \triangleq \mathbb{E}\left[\mathbf{J}_{\text{NR}}(\phi_{l},\phi_{l})\right]$}
Applying expectation operator w.r.t the joint PDF of $\phi_{l}, \psi_{l}$ and $\alpha_{l}$ i.e. $p(\boldsymbol{\theta})$ defined in Eq. (\ref{prior_par_pdf}), we can express Eq. (\ref{eqn_fim_phi_7}) as follows:
\begin{eqnarray}
&&\mathbb{E}\left[\mathbf{J}_{\text{NR}}(\phi_{l},\phi_{l})\right]=\frac{2 \pi^2}{\sigma_{v}^{2}}\mathbb{E}[\alpha_{l}^{2}]\text{tr}\left[\mathbf{K} \left(\widetilde{\mathbf{P}} \otimes \widetilde{\mathbf{Q}}\right)\right]\label{bayesian_fim_phi_eqn_1}\\
&&\text{where}~\widetilde{\mathbf{P}} \triangleq \mathbb{E}[\mathbf{P}]~\text{and}~\widetilde{\mathbf{Q}} \triangleq \mathbb{E}[\mathbf{Q}]~.\label{pqtilde_matrix_def}
\end{eqnarray}
The general elements of matrices $\widetilde{\mathbf{P}}$ and $\widetilde{\mathbf{Q}}$ can be written as below, using the expressions for general elements of matrices $\mathbf{P}$ and $\mathbf{Q}$ in Eq. (\ref{p_matrix_gen_element_def}) and Eq. (\ref{q_matrix_gen_element_def}) respectively:
\begin{eqnarray}
\widetilde{\mathbf{P}}(r,s)&=&\frac{(r-1)(s-1)}{n_{t}}\mathbb{E}\left[e^{-j\pi(s-r)\cos(\phi_{l})} \sin^{2}(\phi_{l})\right]\label{bayesian_fim_phi_eqn_4}\\
%&&Q_{m,n}=\frac{1}{n_{r}}J_{0}\left(\pi(n-m)\right)\label{bayesian_fim_phi_eqn_5}
\widetilde{\mathbf{Q}}(r,s)&=&\frac{1}{n_{r}}\mathbb{E}\left[e^{j \pi(s-r)\cos(\psi_{l})}\right]\label{bayesian_fim_phi_eqn_5}
\end{eqnarray}
Since we assumed a uniform distribution over $[0,\pi)$ for both $\phi_{l}$ and $\psi_{l}$, therefore Eq. (\ref{bayesian_fim_phi_eqn_4}) and Eq. (\ref{bayesian_fim_phi_eqn_5}) can be simplified to the following:
\begin{eqnarray}
\widetilde{\mathbf{P}}(r,s)&=&\frac{(r-1)(s-1)}{n_{t}}\frac{1}{\pi}\int_{0}^{\pi} e^{-j\pi(s-r)\cos(\phi_{l})} \sin^{2}(\phi_{l}) d\phi_{l}\label{bayesian_fim_phi_eqn_6}\\
&=&\frac{(r-1)(s-1)}{2 n_{t}}\left[J_{0}(-\pi(s-r)) + J_{2}(-\pi(s-r))\right]\label{bayesian_fim_phi_eqn_6_1}\\
&=&\frac{1}{n_{r}}\frac{1}{\pi}\int_{0}^{\pi} e^{j \pi(s-r)\cos(\psi_{l})} d\psi_{l} \label{bayesian_fim_phi_eqn_7}\\
\widetilde{\mathbf{Q}}(r,s)&=&\frac{1}{n_{r}}J_{0}\left(\pi(s-r)\right).\label{bayesian_fim_phi_eqn_7_1}
\end{eqnarray}
The Eq. (\ref{bayesian_fim_phi_eqn_6}) follows from Eq. (\ref{bayesian_fim_phi_eqn_6_1}) by using Eq. (\ref{bessel_fn_integral_1}) and Eq. (\ref{bayesian_fim_phi_eqn_7}) follows from Eq. (\ref{bayesian_fim_phi_eqn_7_1}) by using the definition of an $n^{\text{th}}$ order Bessel function of first kind.
\subsection*{Calculation of $\widetilde{\mathbf{J}}_{\text{D}}(\phi_{l},\alpha_{l}) \triangleq \mathbb{E}\left[\mathbf{J}_{\text{NR}}(\phi_{l},\alpha_{l})\right]$}
\begin{eqnarray}
\widetilde{\mathbf{J}}_{\text{D}}(\phi_{l},\alpha_{l})&=&\frac{-2 \pi \mathbb{E}[\alpha_{l}]}{\sigma_{v}^{2}} \text{Im}\left\{\text{tr}\left[\mathbf{K} \left(\widetilde{\mathbf{P}}^{(2)} \otimes \widetilde{\mathbf{Q}}\right) \right]\right\}\label{bayesian_fim_phi_alpha_eqn_1}\\
\text{where}~\widetilde{\mathbf{P}}^{(2)} &\triangleq& \mathbb{E}[\mathbf{P}^{(2)}]~,\label{p2tilde_matrix_def}
\end{eqnarray}
and $\widetilde{\mathbf{Q}}$ is defined as in Eq. (\ref{pqtilde_matrix_def}).

The general element of matrix $\widetilde{\mathbf{P}}^{(2)}$ can be written, using the expression for the general element of $\mathbf{P}$ in Eq. (\ref{p2_general_element_def}), as follows:
\begin{eqnarray}
\widetilde{\mathbf{P}}^{(2)}(r,s)=\frac{(s-1)}{n_{t}}\mathbb{E}\left[e^{-j\pi(s-r)\cos(\phi_{l})} \sin(\phi_{l})\right]\label{bayesian_fim_phi_alpha_eqn_3}
\end{eqnarray}
Since we assumed a uniform distribution over $[0,\pi)$ for $\phi_{l}$, therefore Eq. (\ref{bayesian_fim_phi_alpha_eqn_3}) can be simplified to the following form:
\begin{eqnarray}
\widetilde{\mathbf{P}}^{(2)}(r,s)&=&\frac{(s-1)}{n_{t}}\frac{1}{\pi}\int_{0}^{\pi}e^{-j \pi (s-r)\cos(\phi_{l})} \sin(\phi_{l}) d \phi_{l}\label{bayesian_fim_phi_alpha_eqn_4}\\
&=&\left\{
\begin{array}{l r}
\frac{2(s-1)}{\pi n_{t}} & r=s\\
\frac{(s-1)}{n_{t}}\frac{1}{\pi}\sqrt{\frac{-2}{s-r}}J_{1/2}\left(-\pi (s-r)\right) & r \neq s
\end{array}\right. ~,
\label{bayesian_fim_phi_alpha_eqn_5}
\end{eqnarray}
where Eq. (\ref{bayesian_fim_phi_alpha_eqn_5}) follows from Eq. (\ref{bayesian_fim_phi_alpha_eqn_4}) by using Eq. (\ref{bessel_fn_integral_2}).

%Note that $\widetilde{\mathbf{Q}}$ for the case when $\psi_{l}$ has a uniform distribution over $[0,\pi]$ is given by Eq. (\ref{bayesian_fim_phi_eqn_7}).
\subsection*{Calculation of $\widetilde{\mathbf{J}}_{\text{D}}(\phi_{l},\psi_{l}) \triangleq \mathbb{E}\left[\mathbf{J}_{\text{NR}}(\phi_{l},\psi_{l})\right]$}
\begin{eqnarray}
\widetilde{\mathbf{J}}_{\text{D}}(\phi_{l},\psi_{l})&=&\frac{-2 \pi^{2}\mathbb{E}[\alpha_{l}^{2}]}{\sigma_{v}^{2}}~\text{Re}\left\{\text{tr}\left[\mathbf{K} \left(\widetilde{\mathbf{P}}^{(3)}\otimes \widetilde{\mathbf{Q}}^{(2)}\right)\right]\right\},\label{bayesian_fim_phi_psi_1}\\
&&\text{where}~\widetilde{\mathbf{P}}^{(3)} \triangleq \mathbb{E}\left[\mathbf{P}^{(3)}\right]~\text{and}~\widetilde{\mathbf{Q}}^{(2)} \triangleq \mathbb{E}\left[\mathbf{Q}^{(2)}\right]\label{p3q2tilde_matrix_def}
\end{eqnarray}
The general element in matrix $\widetilde{\mathbf{P}}^{(3)}$ and $\widetilde{\mathbf{Q}}^{(2)}$, using the general element form of matrices $\mathbf{P}^{(3)}$ and $\mathbf{Q}^{(2)}$ given by Eq. (\ref{p3_general_element_def}) and Eq. (\ref{q2_general_element_def}) respectively, can be written as follows:
\begin{eqnarray}
\widetilde{\mathbf{P}}^{(3)}(r,s)&=&\frac{(r-1)}{n_{t}}\mathbb{E}\left[e^{-j\pi(s-r)\cos(\phi_{l})} \sin(\phi_{l})\right]\label{bayesian_fim_phi_psi_eqn_4}\\
\widetilde{\mathbf{Q}}^{(2)}(r,s)&=&\frac{(s-1)}{n_{r}}\mathbb{E}\left[e^{-j\pi(r-s)\cos(\psi_{l})} \sin(\psi_{l})\right]\label{bayesian_fim_phi_psi_eqn_5}
\end{eqnarray}

Since we assumed a uniform distribution over $[0,\pi)$ for both $\phi_{l}$ and $\psi_{l}$,  the Eq. (\ref{bayesian_fim_phi_psi_eqn_4}) and Eq. (\ref{bayesian_fim_phi_psi_eqn_5}) can be simplified as follows:
\begin{eqnarray}
\widetilde{\mathbf{P}}^{(3)}(r,s)&=&\frac{(r-1)}{n_{t}}\frac{1}{\pi}\int_{0}^{\pi}e^{-j \pi (s-r)\cos(\phi_{l})} \sin(\phi_{l}) d \phi_{l}\label{bayesian_fim_phi_psi_eqn_6}\\
&=&\left\{
\begin{array}{lr}
\frac{2(r-1)}{\pi n_{t}} & r=s\\
\frac{(r-1)}{n_{t}}\frac{1}{\pi} \sqrt{\frac{-2}{s-r}} J_{1/2}\left(-\pi (s-r)\right) & r \neq s
\end{array}\right. \label{bayesian_fim_phi_psi_eqn_6_1}\\
\widetilde{\mathbf{Q}}^{(2)}(r,s)&=&\frac{(s-1)}{n_{r}}\frac{1}{\pi}\int_{0}^{\pi}e^{-j \pi (r-s)\cos(\psi_{l})} \sin(\psi_{l}) d \psi_{l}\label{bayesian_fim_phi_psi_eqn_7}\\
&=&\left\{
\begin{array}{lr}
\frac{2(s-1)}{\pi n_{r}} & r=s\\
\frac{(s-1)}{n_{r}}\frac{1}{\pi} \sqrt{\frac{-2}{r-s}} J_{1/2}\left(-\pi (r-s)\right) & r \neq s
\end{array}
\right. \label{bayesian_fim_phi_psi_eqn_7_1}
% &&=\left\{
% \begin{array}{lr}
%  \frac{2(n-1)}{\pi n_{r}} & m=n\\
% \frac{(n-1)}{n_{r}}\frac{1}{\pi} \sqrt{\frac{-2}{m-n}} J_{1/2}\left(-\pi (m-n)\right) & m \neq n
% \end{array}
% \right. \label{bayesian_fim_phi_psi_eqn_7_1}
\end{eqnarray}
The Eq. (\ref{bayesian_fim_phi_psi_eqn_6_1}) follows from Eq. (\ref{bayesian_fim_phi_psi_eqn_6}) and similarly Eq. (\ref{bayesian_fim_phi_psi_eqn_7_1}) follows from Eq. (\ref{bayesian_fim_phi_psi_eqn_7}), by using Eq. (\ref{bessel_fn_integral_2}).
\subsection*{Calculation of $\widetilde{\mathbf{J}}_{\text{D}}(\psi_{l},\alpha_{l}) \triangleq \mathbb{E}\left[\mathbf{J}_{\text{NR}}(\psi_{l},\alpha_{l})\right]$}
\begin{eqnarray}
\widetilde{\mathbf{J}}_{\text{D}}(\psi_{l},\alpha_{l})&=&\frac{2 \pi \mathbb{E}[\alpha_{l}]}{\sigma_{v}^{2}}\text{Im}\left\{\text{tr}\left[\mathbf{K} \left(\widetilde{\mathbf{P}}^{(4)} \otimes \widetilde{\mathbf{Q}}^{(2)}\right)\right]\right\},\label{bayesian_fim_psi_alpha_eqn_1}\\
\text{where}~\widetilde{\mathbf{P}}^{(4)} &\triangleq& \mathbb{E}[\mathbf{P}^{(4)}]~,\label{p4tilde_matrix_def}
\end{eqnarray}
and $\widetilde{\mathbf{Q}}^{(2)}$ is defined as in Eq. (\ref{p3q2tilde_matrix_def}).

Using the expression for general element of matrix $\mathbf{P}^{(4)}$ in Eq. (\ref{p4_general_element_def}), the general element of matrix $\widetilde{\mathbf{P}}^{(4)}$ can be written as follows:
\begin{eqnarray}
\widetilde{\mathbf{P}}^{(4)}(r,s)=\frac{1}{n_{t}}\mathbb{E}\left[e^{j\pi(r-s)\cos(\phi_{l})} \right]\label{bayesian_fim_psi_alpha_eqn_3}
\end{eqnarray}
Since we assumed a uniform distribution, over $[0,\pi)$ for both $\phi_{l}$ and $\psi_{l}$, therefore can simplify Eq. (\ref{bayesian_fim_psi_alpha_eqn_3}) to the expression below:
\begin{eqnarray}
\widetilde{\mathbf{P}}^{(4)}(r,s)&=&\frac{1}{\pi}\int_{0}^{\pi} e^{j\pi(r-s)\cos(\phi_{l})} d\phi_{l}=\frac{J_{0}\left(\pi(r-s)\right)}{n_{t}},\label{bayesian_fim_psi_alpha_eqn_4}
\end{eqnarray}
where the right most term follows from the definition of an $n^{\text{th}}$ order Bessel function of first kind.
\subsection*{Calculation of $\widetilde{\mathbf{J}}_{\text{D}}(\phi_{l},\alpha_{m}) \triangleq \mathbb{E}\left[\mathbf{J}_{\text{NR}}(\phi_{l},\alpha_{m})\right]$}
\begin{eqnarray}
\widetilde{\mathbf{J}}_{\text{D}}(\phi_{l},\alpha_{m})&=&\frac{-2 \pi \mathbb{E}[\alpha_{l}]}{\sigma_{v}^{2}}\text{Im}\left\{\text{tr}\left[\mathbf{K} \left(\widetilde{\mathbf{P}}^{(5)} \otimes \widetilde{\mathbf{Q}}^{(3)}\right)\right]\right\},\label{bayesian_fim_phi_alpha_diff_eqn_1}\\
&&\text{where}~\widetilde{\mathbf{P}}^{(5)} \triangleq \mathbb{E}\left[\mathbf{P}^{(5)}\right]~\text{and}~\widetilde{\mathbf{Q}}^{(3)} \triangleq \mathbb{E}\left[\mathbf{Q}^{(3)}\right]\label{p5q3tilde_matrix_def}
\end{eqnarray}
Using expressions for general element of matrices $\mathbf{P}^{(5)}$ $\mathbf{Q}^{(3)}$ given by Eq. (\ref{p5_general_element_def}) and Eq. (\ref{q3_general_element_def}), the general elements of matrices $\widetilde{\mathbf{P}}^{(5)}$ and $\widetilde{\mathbf{Q}}^{(3)}$ can be written as follows:
\begin{eqnarray}
\widetilde{\mathbf{P}}^{(5)}(r,s)&=&\frac{(s-1)}{n_{t}}\mathbb{E}\left[e^{j\pi(r-1)\cos(\phi_{m})} \right] \mathbb{E}\left[e^{-j\pi(s-1)\cos(\phi_{l})}\sin(\phi_{l})\right]\label{bayesian_fim_phi_alpha_diff_eqn_4}\\
\widetilde{\mathbf{Q}}^{(3)}(r,s)&=&\frac{1}{n_{r}}\mathbb{E}\left[e^{-j\pi(r-1)\cos(\psi_{m})} \right]\mathbb{E}\left[e^{j\pi(s-1)\cos(\psi_{l})} \right]~.\label{bayesian_fim_phi_alpha_diff_eqn_5}
\end{eqnarray}
Since we assumed a uniform distribution, over $[0,\pi)$ for both $\phi_{l}$ and $\psi_{l}$, therefore we can simplify Eq. (\ref{bayesian_fim_phi_alpha_diff_eqn_4}) and Eq. (\ref{bayesian_fim_phi_alpha_diff_eqn_5}) to the following expressions:
\begin{eqnarray}
\widetilde{\mathbf{P}}^{(5)}(r,s)&=&\frac{(s-1)}{n_{t}} J_{0}\left((r-1)\pi\right)\frac{1}{\pi}\int_{0}^{\pi}e^{-j \pi (s-1)\cos(\phi_{l})} \sin(\phi_{l}) d \phi_{l}\label{bayesian_fim_phi_alpha_diff_eqn_6}\\
&=&\frac{(s-1)}{n_{t}} J_{0}\left((r-1)\pi\right) \frac{1}{\pi} \sqrt{\frac{-2}{s-1}} J_{1/2} (-\pi (s-1)) \label{bayesian_fim_phi_alpha_diff_eqn_6_1}\\
\widetilde{\mathbf{Q}}^{(3)}(r,s)&=&\frac{1}{n_{r}} \frac{1}{\pi}\int_{0}^{\pi} e^{-j\pi(r-1)\cos(\psi_{m})} d\psi_{m} \frac{1}{\pi}\int_{0}^{\pi} e^{j\pi(s-1)\cos(\psi_{l})}\label{bayesian_fim_phi_alpha_diff_eqn_7}\\
&=&\frac{1}{n_{r}} J_{0}\left(-(r-1)\pi\right) J_{0}\left((s-1)\pi\right)~.\label{bayesian_fim_phi_alpha_diff_eqn_7_1}
\end{eqnarray}
The Eq. (\ref{bayesian_fim_phi_alpha_diff_eqn_6_1}) follows from Eq. (\ref{bayesian_fim_phi_alpha_diff_eqn_6}) by using Eq. (\ref{bessel_fn_integral_2}), while Eq. (\ref{bayesian_fim_phi_alpha_diff_eqn_7_1}) follows from Eq. (\ref{bayesian_fim_phi_alpha_diff_eqn_7}) by using the definition of an $n^{\text{th}}$ order Bessel function of first kind.
\subsection*{Calculation of $\widetilde{\mathbf{J}}_{\text{D}}(\psi_{l},\alpha_{m}) \triangleq \mathbb{E}\left[\mathbf{J}_{\text{NR}}(\psi_{l},\alpha_{m})\right]$}
\begin{eqnarray}
\widetilde{\mathbf{J}}_{\text{D}}(\psi_{l},\alpha_{m})&=&\frac{2 \pi \mathbb{E}[\alpha_{l}]}{\sigma_{v}^{2}}\text{Im}\left\{\text{tr}\left[\mathbf{K} \left(\widetilde{\mathbf{P}}^{(6)} \otimes \widetilde{\mathbf{Q}}^{(4)}\right)\right]\right\},\label{bayesian_fim_psi_alpha_diff_eqn_1}\\
&&\text{where}~\widetilde{\mathbf{P}}^{(6)} \triangleq \mathbb{E}\left[\mathbf{P}^{(6)}\right]~\text{and}~\widetilde{\mathbf{Q}}^{(4)} \triangleq \mathbb{E}\left[\mathbf{Q}^{(4)}\right]~.\label{p6q4tilde_matrix_def}
\end{eqnarray}
Using expressions for general element of matrices $\mathbf{P}^{(6)}$, $\mathbf{Q}^{(4)}$ given by Eq. (\ref{p6_general_element_def}) and Eq. (\ref{q4_general_element_def}), the general elements of matrices $\widetilde{\mathbf{P}}^{(6)}$ and $\widetilde{\mathbf{Q}}^{(4)}$ can be written as follows:
\begin{eqnarray}
\widetilde{\mathbf{P}}^{(6)}(r,s)&=&\frac{1}{n_{t}}\mathbb{E}\left[e^{j\pi(r-1)\cos(\phi_{m})} \right]\mathbb{E}\left[e^{-j\pi(s-1)\cos(\phi_{l})} \right]\label{bayesian_fim_psi_alpha_diff_eqn_4}\\
\widetilde{\mathbf{Q}}^{(4)}(r,s)&=&\frac{(s-1)}{n_{r}}\mathbb{E}\left[e^{-j\pi(r-1)\cos(\psi_{m})} \right]\mathbb{E}\left[e^{j\pi(s-1)\cos(\psi_{l})} \sin(\psi_{l})\right]~.\label{bayesian_fim_psi_alpha_diff_eqn_5} 
\end{eqnarray}

Since we assumed $\phi_{l},~\phi_{m},~\psi_{l}$ and $\psi_{m}$ to be uniformly distributed over $[0,\pi)$, therefore we can simplify Eq. (\ref{bayesian_fim_psi_alpha_diff_eqn_4}) and Eq. (\ref{bayesian_fim_psi_alpha_diff_eqn_5}) to the following expressions:
\begin{eqnarray}
\widetilde{\mathbf{P}}^{(6)}(r,s)&=&\frac{1}{n_{t}} \frac{1}{\pi} \int_{0}^{\pi} e^{j\pi(r-1)\cos(\phi_{m})} d\phi_{m} \frac{1}{\pi} \int_{0}^{\pi} e^{-j\pi(s-1)\cos(\phi_{l})} d\phi_{l}\label{bayesian_fim_psi_alpha_diff_eqn_6}\\
&=&\frac{1}{n_{t}}J_{0}\left((r-1)\pi\right)J_{0}\left(-(s-1)\pi\right)\label{bayesian_fim_psi_alpha_diff_eqn_6_1}\\
%&=&\frac{1}{n_{t}}J_{0}\left((r-1)\pi\right)J_{0}^{*}\left(-(s-1)\pi\right) \label{bayesian_fim_psi_alpha_diff_eqn_6}\\
\widetilde{\mathbf{Q}}^{(4)}(r,s)&=& \left(\frac{(s-1)}{n_{r}}\frac{1}{\pi} \int_{0}^{\pi} e^{-j\pi(r-1)\cos(\psi_{m})} d\psi_{m}\right) \left(\frac{1}{\pi} \int_{0}^{\pi} e^{j\pi(s-1)\cos(\psi_{l})} \sin(\psi_{l}) d \psi_{l}\right)\label{bayesian_fim_psi_alpha_diff_eqn_7}\\
%&=&\frac{(s-1)}{n_{r}}J_{0}\left(-(r-1)\pi\right) \frac{1}{\pi}\int_{0}^{\pi}e^{j \pi (s-1) \cos(\psi_{l})} \sin(\psi_{l}) d \psi_{l} \nonumber\\
&=&\left(\frac{(s-1)}{n_{r}}J_{0}\left(-(r-1)\pi\right)\right) \left(\frac{1}{\pi} \sqrt{\frac{2}{s-1}} J_{1/2}\left(\pi (s-1)\right)\right)~.\label{bayesian_fim_psi_alpha_diff_eqn_7_1}
\end{eqnarray}
The Eq. (\ref{bayesian_fim_psi_alpha_diff_eqn_6_1}) follows from Eq. (\ref{bayesian_fim_psi_alpha_diff_eqn_6}) by using the definition of an $n^{\text{th}}$ order Bessel function of first kind. The Eq. (\ref{bayesian_fim_psi_alpha_diff_eqn_7_1}) follows from Eq. (\ref{bayesian_fim_psi_alpha_diff_eqn_7}) by using Eq. (\ref{bessel_fn_integral_2}).
\subsection*{Calculation of $\widetilde{\mathbf{J}}_{\text{D}}(\phi_{l},\phi_{m}) \triangleq \mathbb{E}\left[\mathbf{J}_{\text{NR}}(\phi_{l},\phi_{m})\right]$}
\begin{eqnarray}
\widetilde{\mathbf{J}}_{\text{D}}(\phi_{l},\phi_{m})&=&\frac{2 \pi^{2} \mathbb{E}[\alpha_{l}] \mathbb{E}[\alpha_{m}]}{\sigma_{v}^{2}}\text{Re}\left\{\text{tr}\left[\mathbf{K} \left(\widetilde{\mathbf{P}}^{(7)} \otimes \widetilde{\mathbf{Q}}^{(3)}\right)\right]\right\},\label{bayesian_fim_phi_phi_diff_eqn_1}\\
\text{where}~\widetilde{\mathbf{P}}^{(7)} &\triangleq& \mathbb{E}[\mathbf{P}^{(7)}]~,\label{p7tilde_matrix_def}
\end{eqnarray}
and $\widetilde{\mathbf{Q}}^{(3)}$ is defined as in Eq. (\ref{p5q3tilde_matrix_def}).

The general element of matrix $\widetilde{\mathbf{P}}^{(7)}$ can be expressed using the expression for general element of matrix $\mathbf{P}^{(7)}$ given by Eq. (\ref{p7_general_element_def}) and simplified further by noting that $\phi_{l}, ~ \phi_{m}$ to be independent and uniformly distributed over $[0,\pi)$:
\begin{eqnarray}
\widetilde{\mathbf{P}}^{(7)}(r,s)&=&\frac{(r-1)(s-1)}{n_{t}} \mathbb{E}\left[e^{j \pi (r-1) \cos(\phi_{m})} \sin(\phi_{m})\right] \mathbb{E}\left[e^{-j \pi (s-1) \cos(\phi_{l})} \sin(\phi_{l})\right] \label{bayesian_fim_phi_phi_diff_eqn_3}\\
&=&\frac{(r-1)(s-1)}{n_{t}} \frac{1}{\pi} \int_{0}^{\pi} e^{j \pi (r-1) \cos(\phi_{m})} \sin(\phi_{m}) d \phi_{m} \frac{1}{\pi} \int_{0}^{\pi} e^{-j \pi (s-1) \cos(\phi_{l})} \sin(\phi_{l}) d \phi_{l} \nonumber\\
&=&\left(\frac{(r-1)(s-1)}{n_{t}} \frac{1}{\pi} \sqrt{\frac{2}{r-1}} J_{1/2}(\pi (r-1))\right) \left(\frac{1}{\pi} \sqrt{\frac{-2}{s-1}} J_{1/2}(-\pi (s-1))\right)~. \label{bayesian_fim_phi_phi_diff_eqn_4_1}
\end{eqnarray}
The Eq. (\ref{bayesian_fim_phi_phi_diff_eqn_4_1}) follow from the equation above it by using Eq. (\ref{bessel_fn_integral_2}).
\subsection*{Calculation of $\widetilde{\mathbf{J}}_{\text{D}}(\psi_{l},\psi_{m}) \triangleq \mathbb{E}\left[\mathbf{J}_{\text{NR}}(\psi_{l},\psi_{m})\right]$}
\begin{eqnarray}
\widetilde{\mathbf{J}}_{\text{D}}(\psi_{l},\psi_{m})&=&\frac{2 \pi^{2} \mathbb{E}[\alpha_{l}] \mathbb{E}[\alpha_{m}]}{\sigma_{v}^{2}}\text{Re}\left\{\text{tr}\left[\mathbf{K} \left(\widetilde{\mathbf{P}}^{(8)} \otimes \widetilde{\mathbf{Q}}^{(5)}\right)\right]\right\},\label{bayesian_fim_psi_psi_diff_eqn_1}\\
&&\text{where}~\widetilde{\mathbf{P}}^{(8)} \triangleq \mathbb{E}\left[\mathbf{P}^{(8)}\right]~\text{and}~\widetilde{\mathbf{Q}}^{(5)} \triangleq \mathbb{E}\left[\mathbf{Q}^{(5)}\right]\label{p8q5tilde_matrix_def}
\end{eqnarray}
Using expressions for general element of matrices $\mathbf{P}^{(8)}$, $\mathbf{Q}^{(5)}$ given by Eq. (\ref{p8_general_element_def}) and Eq. (\ref{q5_general_element_def}), the general elements of matrices $\widetilde{\mathbf{P}}^{(8)}$ and $\widetilde{\mathbf{Q}}^{(5)}$ can be written as follows:
\begin{eqnarray}
\widetilde{\mathbf{P}}^{(8)}(r,s)&=&\frac{1}{n_{t}} \mathbb{E}\left[e^{j \pi (r-1) \cos(\phi_{l})}\right] \mathbb{E}\left[e^{-j \pi (s-1) \cos(\phi_{m})}\right] \label{bayesian_fim_psi_psi_diff_eqn_4}\\
\widetilde{\mathbf{Q}}^{(5)}(r,s)&=&\frac{(r-1)(s-1)}{n_{r}} \mathbb{E}\left[e^{-j \pi (r-1) \cos(\psi_{l})} \sin(\psi_{l})\right] \mathbb{E}\left[e^{j \pi (s-1) \cos(\psi_{m})} \sin(\psi_{m})\right] \label{bayesian_fim_psi_psi_diff_eqn_5}
\end{eqnarray}
Since we assumed $\phi_{l}, ~ \phi_{m}, ~\psi_{l}$ and $\psi_{m}$ to be independent of each other and are uniform distributed over $[0,\pi)$, therefore we can simplify Eq. (\ref{bayesian_fim_psi_psi_diff_eqn_4}) and Eq. (\ref{bayesian_fim_psi_psi_diff_eqn_5}) to the following:
\begin{eqnarray}
\widetilde{\mathbf{P}}^{(8)}(r,s)&=&\left(\frac{1}{n_{t}} \frac{1}{\pi} \int_{0}^{\pi} e^{j \pi (r-1) \cos(\phi_{l})} d\phi_{l}\right) \left(\frac{1}{\pi} \int_{0}^{\pi} e^{-j \pi (s-1) \cos(\phi_{m})} d\phi_{m}\right)\nonumber\\
&=&\frac{1}{n_{t}} J_{0}\left(\pi(r-1)\right) J_{0}\left(-\pi(s-1)\right)\label{bayesian_fim_psi_psi_diff_eqn_6}\\
\widetilde{\mathbf{Q}}^{(5)}(r,s)&=&\frac{(r-1)(s-1)}{n_{r}} \frac{1}{\pi} \int_{0}^{\pi} e^{-j \pi (r-1) \cos(\psi_{l})} \sin(\psi_{l}) d \psi_{l} \frac{1}{\pi} \int_{0}^{\pi} e^{j \pi (s-1) \cos(\psi_{m})} \sin(\psi_{m}) d \psi_{m}  \nonumber\\
&=&\frac{(r-1)(s-1)}{n_{r}} \frac{1}{\pi} \sqrt{\frac{-2}{r-1}} J_{1/2}\left(-\pi(r-1)\right) \frac{1}{\pi} \sqrt{\frac{2}{s-1}} J_{1/2}\left(\pi(s-1)\right)~. \label{bayesian_fim_psi_psi_diff_eqn_7_1}
\end{eqnarray}
The Eq. (\ref{bayesian_fim_psi_psi_diff_eqn_6}) follows from the equation above it by a straightforward application of definition of an $n^{\text{th}}$ order Bessel function of first kind. The Eq. (\ref{bayesian_fim_psi_psi_diff_eqn_7_1}) follows from the equation above it via application of Eq. (\ref{bessel_fn_integral_2}).
\subsection*{Calculation of $\widetilde{\mathbf{J}}_{\text{D}}(\psi_{l},\psi_{l}) \triangleq \mathbb{E}\left[\mathbf{J}_{\text{NR}}(\psi_{l},\psi_{l})\right]$}
\begin{eqnarray}
\widetilde{\mathbf{J}}_{\text{D}}(\psi_{l},\psi_{l})&=&\frac{2 \pi^2 \mathbb{E}[\alpha_{l}^{2}]}{\sigma_{v}^{2}}\text{tr}\left(\mathbf{K} \left(\widetilde{\mathbf{P}}^{(4)} \otimes \widetilde{\mathbf{Q}}^{(6)}\right) \right)\label{bayesian_fim_psi_eqn_1}\\
\text{where}~\widetilde{\mathbf{Q}}^{(6)} &\triangleq& \mathbb{E}[\mathbf{Q}^{(6)}]~,\label{q6tilde_matrix_def}
\end{eqnarray}
and $\widetilde{\mathbf{P}}^{(4)}$ is defined as in Eq. (\ref{p4tilde_matrix_def}).

Using the expression for general element of $\mathbf{Q}^{(6)}$ in Eq. (\ref{q6_general_element_def}) we can express the general element of $\widetilde{\mathbf{Q}}^{(6)}$ to be the following:
\begin{eqnarray}
\widetilde{\mathbf{Q}}^{(6)}(r,s)&=&\frac{(r-1)(s-1)}{n_{r}}\mathbb{E}\left[e^{j\pi(s-r)\cos(\psi_{l})} \sin^{2}(\psi_{l})\right]~.\label{bayesian_fim_psi_eqn_4}
\end{eqnarray}
If we assume a uniform distribution over $[0,\pi)$ for $\psi_{l}$, then Eq. (\ref{bayesian_fim_psi_eqn_4}) can be simplified to the following:
\begin{eqnarray}
\widetilde{\mathbf{Q}}^{(6)}(r,s)&=&\frac{(r-1)(s-1)}{n_{r}} \frac{1}{\pi}\int_{0}^{\pi}e^{j\pi(s-r)\cos(\psi_{l})} \sin^{2}(\psi_{l})~,\nonumber\\
&=&\frac{(r-1)(s-1)}{2 n_{r}}\left[J_{0}(\pi(s-r)) + J_{2}(\pi(s-r))\right]~.\label{bayesian_fim_psi_eqn_5}
\end{eqnarray}
The Eq. (\ref{bayesian_fim_psi_eqn_5}) follows from the equation above it via a straightforward application of Eq. (\ref{bessel_fn_integral_1}).
\subsection*{Calculation of $\widetilde{\mathbf{J}}_{\text{D}}(\phi_{l},\psi_{m}) \triangleq \mathbb{E}\left[\mathbf{J}_{\text{NR}}(\phi_{l},\psi_{m})\right]$}
\begin{eqnarray}
\widetilde{\mathbf{J}}_{\text{D}}(\phi_{l},\psi_{m})&=&\frac{-2 \pi^{2}\mathbb{E}[\alpha_{l}] \mathbb{E}[\alpha_{m}]}{\sigma_{v}^{2}}\text{Re}\left\{\text{tr}\left[\mathbf{K} \left(\widetilde{\mathbf{P}}^{(9)} \otimes \widetilde{\mathbf{Q}}^{(7)}\right)\right]\right\},\label{bayesian_fim_phi_psi_diff_eqn_1}\\
&&\text{where}~\widetilde{\mathbf{P}}^{(9)} \triangleq \mathbb{E}\left[\mathbf{P}^{(9)}\right]~\text{and}~\widetilde{\mathbf{Q}}^{(7)} \triangleq \mathbb{E}\left[\mathbf{Q}^{(7)}\right]\label{p9q7tilde_matrix_def}
\end{eqnarray}
Using expressions for general element of matrices $\mathbf{P}^{(9)}$ $\mathbf{Q}^{(7)}$ given by Eq. (\ref{p9_general_element_def}) and Eq. (\ref{q7_general_element_def}), the general elements of matrices $\widetilde{\mathbf{P}}^{(9)}$ and $\widetilde{\mathbf{Q}}^{(7)}$ can be written as follows:
\begin{eqnarray}
\widetilde{\mathbf{P}}^{(9)}(r,s)&=&\frac{(r-1)}{n_{t}} \mathbb{E}\left[e^{j \pi (r-1) \cos(\phi_{l})} \sin(\phi_{l})\right] \mathbb{E}\left[e^{-j \pi (s-1) \cos(\phi_{m})}\right] \label{bayesian_fim_phi_psi_diff_eqn_4}\\
\widetilde{\mathbf{Q}}^{(7)}(r,s)&=&\frac{(s-1)}{n_{r}} \mathbb{E}\left[e^{-j \pi (r-1) \cos(\psi_{l})}\right] \mathbb{E}\left[e^{j \pi (s-1) \cos(\psi_{m})} \sin(\psi_{m})\right] \label{bayesian_fim_phi_psi_diff_eqn_5}
\end{eqnarray}
Since we assumed $\phi_{l}, ~ \phi_{m}, ~ \psi_{l}$ and $\psi_{m}$, to be of independent of each other and take a uniform distribution over $[0,\pi)$, therefore we can simplify Eq. (\ref{bayesian_fim_phi_psi_diff_eqn_4}) and Eq. (\ref{bayesian_fim_phi_psi_diff_eqn_5}) to the following:
\begin{eqnarray}
\widetilde{\mathbf{P}}^{(9)}(r,s)&=&\frac{(r-1)}{n_{t}} \frac{1}{\pi} \int_{0}^{\pi} e^{-j \pi (s-1) \cos(\phi_{m})} d\phi_{m}  \frac{1}{\pi} \int_{0}^{\pi} e^{j(r-1)\pi \cos(\phi_{l})} \sin(\phi_{l}) d \phi_{l} \label{bayesian_fim_phi_psi_diff_eqn_6}\\
&=&\frac{(r-1)}{n_{t}} J_{0}\left(-\pi(s-1)\right) \frac{1}{\pi}\sqrt{\frac{2}{r-1}}J_{1/2}((r-1)\pi)\label{bayesian_fim_phi_psi_diff_eqn_6_1}\\
\widetilde{\mathbf{Q}}^{(7)}(r,s)&=&\frac{(s-1)}{n_{r}} \frac{1}{\pi} \int_{0}^{\pi} e^{-j \pi (r-1) \cos(\psi_{l})} d\psi_{l}  \frac{1}{\pi} \int_{0}^{\pi} e^{j \pi (s-1) \cos(\psi_{m})} \sin(\psi_{m}) d \psi_{m} \label{bayesian_fim_phi_psi_diff_eqn_7}\\
&=&\frac{(s-1)}{n_{r}} J_{0}\left(-\pi(r-1)\right) \frac{1}{\pi} \sqrt{\frac{2}{s-1}}J_{1/2}((s-1)\pi)~. \label{bayesian_fim_phi_psi_diff_eqn_7_1}
\end{eqnarray}
Equations (\ref{bayesian_fim_phi_psi_diff_eqn_6_1}) and (\ref{bayesian_fim_phi_psi_diff_eqn_7_1}) follow from Eq. (\ref{bayesian_fim_phi_psi_diff_eqn_6}) and Eq. (\ref{bayesian_fim_phi_psi_diff_eqn_7}) respectively, by using the definition of an $n^{\text{th}}$ order Bessel function of first kind and the Eq. (\ref{bessel_fn_integral_2}).
\subsection*{Calculation of $\widetilde{\mathbf{J}}_{\text{D}}(\alpha_{l},\alpha_{l}) \triangleq \mathbb{E}\left[\mathbf{J}_{\text{NR}}(\alpha_{l},\alpha_{l})\right]$}
\begin{eqnarray}
\widetilde{\mathbf{J}}_{\text{D}}(\alpha_{l},\alpha_{l})&=&\frac{2}{\sigma_{v}^{2}}\text{tr}\left[\mathbf{K} \left(\widetilde{\mathbf{P}}^{(4)} \otimes \widetilde{\mathbf{Q}}\right)\right],\label{bayesian_fim_alpha_eqn_1}
\end{eqnarray}
where $\widetilde{\mathbf{Q}}$ and $\widetilde{\mathbf{P}}^{(4)}$ are defined as in Eq. (\ref{pqtilde_matrix_def}) and Eq. (\ref{p4tilde_matrix_def}).
\subsection*{Calculation of $\widetilde{\mathbf{J}}_{\text{D}}(\alpha_{l},\alpha_{m}) \triangleq \mathbb{E}\left[\mathbf{J}_{\text{NR}}(\alpha_{l},\alpha_{m})\right]$}
\begin{eqnarray}
\widetilde{\mathbf{J}}_{\text{D}}(\alpha_{l},\alpha_{m})&=&\frac{2}{\sigma_{v}^{2}}\text{Re}\left\{\text{tr}\left[\mathbf{K} \left(\widetilde{\mathbf{P}}^{(6)} \otimes \widetilde{\mathbf{Q}}^{(3)}\right)\right]\right\},\label{bayesian_fim_alpha_alpha_diff_eqn_1}
\end{eqnarray}
where $\widetilde{\mathbf{Q}}^{(3)}$ and $\widetilde{\mathbf{P}}^{(6)}$ are defined as in Eq. (\ref{p5q3tilde_matrix_def}) and Eq. (\ref{p6q4tilde_matrix_def}).
\bibliographystyle{IEEEbib}
\bibliography{lax_mmwave.bib}
\end{document}